\documentclass[twoside]{book}

\usepackage{multido}
\makeatletter
\let\FPaddMultiDo\FPadd
\let\FPsubMultiDo\FPsub
\def\multido@init@n#1#2#3{
  \edef#3{#1}

  \ifnum\multido@count<\z@
    \expandafter\FPsubMultiDo%
  \else\expandafter
    \FPaddMultiDo%
  \fi
  {0}{#2}\multido@temp
  \multido@addtostep{\do\FPaddMultiDo{\do#3}{\multido@temp}{\do#3}}}
\makeatother
\usepackage{fp}

\makeatletter
\def\magnification{\afterassignment\m@g\count@}
\def\m@g{\mag\count@}
\newlength{\myVSpace}
\usepackage[left=3cm,right=3cm]{geometry}
\usepackage{stmaryrd}
\usepackage{amsmath, amssymb}
\usepackage[french]{babel}
\usepackage[latin1]{inputenc}
\usepackage{color}
\usepackage{epsfig}
\usepackage{graphicx}
\usepackage{mathrsfs}
\usepackage{geometry}
\usepackage{amsmath,amsfonts,amssymb, amsthm}
\renewcommand{\d}{\mbox{d}}
\usepackage{pstricks,pst-plot}
\usepackage{wrapfig}
\renewcommand{\d}{\mbox{d}}
\renewcommand{\det}{\mbox{det}}
\newcommand{\E}{\mathbb{E}}
\renewcommand{\H}{\mathbb{H}}
\newcommand{\R}{\mathbb{R}}
\newcommand{\N}{\mathbb{N}}
\newcommand{\Q}{\mathbb{Q}}
\newcommand{\G}{\mathbb{G}}
\newcommand{\M}{\mathbb{M}}
\renewenvironment{proof}{
{\par\sc Proof.}}{\qed\par\smallskip}

\newcommand{\p}[1]{\overset{.}{#1}}
\newcommand{\pp}[1]{\overset{..}{#1}}
\renewcommand{\E}[1]{\mathbb{E}\left(#1\right)}

\newcommand{\sind}{\mbox{sind}}
\newcommand{\angleplat}[7]{
\FPupn\xU{#1}
\FPupn\yU{#2}
\FPupn\xZ{#3}
\FPupn\yZ{#4}
\FPupn\xD{#5}
\FPupn\yD{#6}
\FPupn\L{#7}
\FPsub\VxU{\xU}{\xZ}
\FPsub\VyU{\yU}{\yZ}
\FPsub\VxD{\xD}{\xZ}
\FPsub\VyD{\yD}{\yZ}
\FPabs\VxUa\VxU
\FPabs\VyUa\VyU
\FPabs\VxDa\VxD
\FPabs\VyDa\VyD
\FPpow\carxU{\VxUa}{2}
\FPpow\caryU{\VyUa}{2}
\FPpow\carxD{\VxDa}{2}
\FPpow\caryD{\VyDa}{2}
\FPadd\carU{\carxU}{\caryU}
\FPadd\carD{\carxD}{\caryD}
\FProot\lU{\carU}{2}
\FProot\lD{\carD}{2}
\FPdiv\rU{\L}{\lU}
\FPdiv\rD{\L}{\lD}
\FPmul\VxUn\VxU\rU
\FPmul\VyUn\VyU\rU
\FPmul\VxDe\VxD\rD
\FPmul\VyDe\VyD\rD
\FPadd\xUn\xZ\VxUn
\FPadd\yUn\yZ\VyUn
\FPadd\xDe\xZ\VxDe
\FPadd\yDe\yZ\VyDe
\FPadd\xT\xUn\VxDe
\FPadd\yT\yUn\VyDe
\psline(\xUn,\yUn)(\xT,\yT)(\xDe,\yDe) }

\makeatletter
\renewcommand\theequation{\arabic{equation}}
\@addtoreset{equation}{chapter} \makeatother

\makeatletter
\def\@makechapterhead#1{%
  \vspace*{30\p@}%
  {\parindent \z@ \raggedright \normalfont
    \interlinepenalty\@M
    \ifnum \c@secnumdepth >\m@ne
    \fi
    \centering\Huge #1\par\nobreak
    \vskip 40\p@
  }
  }
\def\@makeschapterhead#1{%
  \vspace*{30\p@}%
  {\parindent \z@ \raggedright
    \normalfont
    \interlinepenalty\@M
    \centering\Huge#1\par\nobreak
    \vskip 40\p@
  }}

\makeatother

\begin{document}
\renewcommand{\contentsname}{Table of contents}

\setcounter{tocdepth}{2}

\thispagestyle{empty}

\vskip 4cm

\begin{center}
{\Huge Heat kernel expansion for a family of stochastic volatility
models : {\rm\Huge{$\delta$}}-geometry }
\end{center}
\vskip 2cm

{{\bf Paul Bourgade, Olivier Croissant}} \\

\noindent
{Ixis-CIB, fixed income quantitative research\footnote{The views herein are the authors' ones and do not
necessarily reflect those of Ixis-cib.}}\\
(e-mail: pbourgade@ixis-cib.com, ocroissant@ixis-cib.com)\\
\vspace{3cm}

{\bf Abstract.}
In this paper, we study a family of stochastic volatility processes;
this family features a mean reversion term for the volatility and a double CEV-like exponent that generalizes
SABR and Heston's models.
We derive approximated closed form formulas for the digital prices, the local and implied volatilities.
Our formulas are efficient for small maturities.

Our method is based on differential geometry,
especially small time diffusions on riemanian spaces. This geometrical point of view can be extended to other processes,
and is very accurate to produce variate smiles for small maturities and small moneyness.

\vspace{1cm}


\vspace{0.5cm}

{\bf Mathematics subjects classification :} 58J65

\vspace{0.5cm}


\newpage

\thispagestyle{empty}

\vskip 1cm

\begingroup

\footnotesize
\parindent0pt
\rightskip-5pt

\noindent{\it References}
\medskip
\rm

\leftskip 2em
\noindent
\llap{1.} {\sc Berestycki Henri, Busca Jérôme and Florent Igor,} {\it Computing the Implied Volatility},
Communications on Pure and Applied Mathematics, Vol. LVII, 0001-0022 (2004).\\
\llap{2.} {\sc Bost Jean Benoit,} {\it Courbes elliptiques et formes modulaires,} lecture at \'Ecole Polytechnique.\\
\llap{3.} {\sc Bourguignon Jean Pierre and Deruelle Nathalie,} {\it Relativité générale,} lecture at \'Ecole Polytechnique.\\
\llap{4.} {\sc Carr Peter and Madan Dilin,} {\it Towards a theory of volatility trading,} Volatility, ed. R.A. Jarrow, Risk Publications.\\
\llap{5.} {\sc El Karoui Nicole,} {\it Modèles stochastiques en finance : finance,} lecture at \'Ecole Polytechnique.\\
\llap{6.} {\sc El Karoui Nicole and Gobet Emmanuel,} {\it Modèles stochastiques en finance : introduction au calcul stochastique,} lecture at \'Ecole Polytechnique.\\
\llap{7.} {\sc Gatheral Jim,} {\it Stochastic Volatility and Local Volatility}, Case Studies in Financial Modelling Course Notes,
Courant Institute of Mathematical Sciences, Fall Term, 2004.\\
\llap{8.} {\sc Hagan Patrick S., Kumar Deep, Lesniewski Andrew S. and Woodward Diana E.}, {\it Managing Smile Risk}.\\
\llap{9.} {\sc Hagan Patrick, Lesniewski Andrew, Woodward Diana}, {\it Probability Distribution in the SABR Model of
Stochastic Volatility}, preprint.\\
\llap{10.} {\sc Hagan Patrick S.  and Woodward Diana E.,} {\it Equivalent Black Volatilities},
Applied Mathematical Finance 6, 147-157 (1999).\\
\llap{11.} {\sc Henry-Labordère Pierre,} {\it A General Asymptotic Implied Volatility for Stochastic Volatility Models,}
preprint.\\
\llap{12.} {\sc Hsu Elton P.,} {\it Stochastic Analysis on Manifolds,} American Mathematical Society vol. 38,
Graduate Studies in Mathematics, 2001.\\
\llap{13.} {\sc Jost Jûrgen,} {\it Riemannian Geometry and Geometric Analysis,} Springer Universitext, 2002.\\
\llap{14.} {\sc Karatzas Ioannis , Shreve Steven E.,} {\it Brownian Motion and Stochastic Calculus,} Springer.\\
\llap{15.} {\sc Lang Serge,} {\it Fundamentals of Differential Geometry,} Springer.\\
\llap{16.} {\sc Milnor J.,} {\it Morse Theory,} Annals of Math. Studies, Princeton University Press (1963).\\
\llap{17.} {\sc Minakshisundaram and Pleijel},
{\it Some properties of the eigenfunctions of the Laplace-operator on Riemannian manifolds},
Canadian J. Math. 1 (1949), 242-256.\\
\llap{18.} {\sc Molchanov S. A.}, {\it Diffusion Processes and Riemannian Geometry}, Russian Math. Surveys 30 : 1 (1975).\\
\llap{19.} {\sc Pansu Pierre,} {\it Cours de géométrie différentielle,} département de mathématiques d'Orsay.\\
\llap{20.} {\sc Tyner David R.,} {\it Applications of variational methods in Riemannian geometry,} department of mathematics and statistics,
Queen's University, Kingston.\\

\endgroup

\renewcommand{\contentsname}{Table of contents}

\tableofcontents

\it

\chapter*{Introduction}
\addcontentsline{toc}{chapter}{Introduction}

In this note, we consider the following process\footnote{In the
first chapter, with generalities about stochastic volatility
problems, we will understand why we consider this particular
process} with stochastic volatility : \rm
\begin{equation}
\left\{
\begin{array}{cccccc}
\d F(t)&=&&&\sigma(t)^\delta c(F(t))&\d W(t)\\
\d \sigma(t)&=&\lambda'(\mu'-\sigma(t))\d t&+&\nu\sigma(t)^\delta&\d Z(t)\\
\E{\d W(t) \d Z(t)}&=&\rho \d t&&&
\end{array}
\right.
.
\end{equation}
\it

\begin{itemize}
\item for $\delta=1$, $c(F)=F^\beta$ and $\lambda'=0$ we get the SABR model;
\item for $\delta=1/2$ and $c(F)=F$ we get Heston's model.
\end{itemize}

A direct application of Feller's criterion\footnote{see {\sc Ioannis Karatzas, Steven E. Shreve,} {\it Brownian Motion and Stochastic Calculus,} Springer.},
detailed in appendix 1, shows that for $1/2<\delta\leqslant 1$
there is no explosion for the stochastic volatility, that is to say
$$P\left[(\forall t\geqslant 0)\ (\sigma(t)\in]0,+\infty[)\right]=1.$$
This result remains true for $\delta=1/2$ if we add the condition $2\lambda \mu/\nu^2\geqslant 1$.
We suppose in the following that we are in such cases.\\

Let us write $G_{F,\Sigma}(\tau,f,\sigma)$ be the density of probability to get to the point $(F,\Sigma)$,
leaving from the point $(f,\sigma)$ after a time $\tau$. Then $G$ follows the classical differential
equation (where $\lambda'$ and $\mu'$ should be adjusted\footnote{{\sc Jim Gatheral,} {\it Stochastic Volatility and
Local Volatility},
Case Studies in Financial Modelling Course Notes,
Courant Institute of Mathematical Sciences,
Fall Term, 2004.})

\begin{equation}
\frac{\partial G}{\partial \tau}=\frac{1}{2}\sigma^{2\delta}\left[c(f)^2\frac{\partial^2G}{\partial f^2}+2\rho \nu c(f)\frac{\partial^2G}{\partial \sigma\partial f}+\nu^2\frac{\partial^2 G}
{\partial \sigma^2}\right]+\lambda'(\sigma-\mu')\frac{\partial G}{\partial \sigma}.
\end{equation}

The goal for this note is to get suitable approximations for the $\delta$-model, for small $\tau$.
There are three steps to get them.

\begin{itemize}

\item We first explain how to convert our stochastic problem into a {\bf geometrical} one. We introduce the Laplace Beltrami operator
to have a geometric point of view, and then use a suitable isometry to work in a simpler space, called the $\delta$-space.

\item Thanks to this geometric point of view, we then give a {\bf first-order asymptotics for the probability density $G$}.
Moreover, we want this expression to be computed for all $F$ and $\delta$ without having to calculate any integral.
The main tool for this part is Molchanov's theorem\footnote{{\sc S. A. Molchanov}, {\it Diffusion Processes and Riemannian Geometry}, Russian Math. Surveys 30 : 1 (1975).}.

\item The third one is to get a first order (in $\tau$) approximation for :
\begin{itemize}
\item the {\bf transition probability} from $(\sigma,f)$ to $F$, where we integrate over $\Sigma$ because the final volatility is not observed :
$$P_F(\tau,\sigma,f)=\int_\R G_{F,\Sigma}(\tau,f,\sigma)\d \Sigma.$$
This gives us the price for digital options.

\item the {\bf implied volatility} $\sigma_{BS}(K)$, defined by the equality of the prices of calls
for (1) and a Black-Scholes model with volatility $\sigma_{BS}(K)$ :
$$\operatorname{Call_{BS}}(\sigma_{BS}(K),K)=\operatorname{Call}_{\delta-model}(\sigma,K).$$
Actually, we first give an approximation for the local volatility (defined by $\sigma_K(T,f,\sigma)^2\d T=\E{\left(\d F(T)\right)^2\mid F(T)=K}$)
and then use Hagan's loc\slash imp formula\footnote{{\sc Patrick S. Hagan and Diana E. Woodward,} {\it Equivalent Black Volatilities},
Applied Mathematical Finance 6, 147-157 (1999).}.
\end{itemize}

\end{itemize}

Such approximations are given in literature for the SABR model, that is to say $\delta=1$. The most successful closed form formula giving
an approximation for options is Hagan's\footnote{{\sc Patrick S. Hagan, Deep Kumar, Andrew S. Lesniewski and Diana E. Woodward}, {\it Managing Smile Risk}.}.
In a preprint\footnote{{\sc Patrick Hagan, Andrew Lesniewski, Diana Woodward}, {\it Probability Distribution in the SABR Model of
Stochastic Volatility}.} Andrew Lesniewski uses a geometrical method (in the hyperbolic plane) to get the density's asymptotics, but he uses a
Dyson formula instead of Molchanov's theorem. Consequently our results are sensibly different.

Berestycki\footnote{{\sc Henri Berestycki,
Jérôme Busca and
Igor Florent,} {\it Computing the Implied Volatility}, Communications on Pure and Applied Mathematics, Vol. LVII, 0001-0022 (2004)
\copyright 2004 Wiley Periodicals, Inc.} gives an evolution equation for the smile. His PDE is generally difficult to solve
and hides the geometrical aspect of the problem. This is the reason why, for our $\delta$-model, we do not try to use the available PDE and directly calculate the first order.\\

\rm

\chapter{A geometrical view of stochastic volatility problems}{}{}

We consider the following general process,
$$
\left\{
\begin{array}{ccccc}
\d F&=&f_1(\Sigma,F)\d t &+&f_2(\Sigma,F)\d W_1\\
\d \Sigma&=&g_1(\Sigma)\d t &+&g_2(\Sigma)\d W_2\\
\end{array}
\right.,
$$
with $<\d W_1,\d W_2>=\rho \d t$. We call $G$ the transition
probability density, from a state $(f_0,\sigma_0)$ to a state
$(f,\sigma)$.
Then the backward Fokker Planck equation can be written
$$
\partial_\tau G=\frac{1}{2}\left(f_2^2\partial_{ff}G+g_2^2\partial_{\sigma\sigma}G+2\rho f_2 g_2\partial_{\sigma f}G\right)
+f_1\partial_f G+g_1 \partial_\sigma G.
$$

The Laplace Beltrami operator is defined as
$$\Delta^\G=g^{-1/2}\partial_\mu\left(g^{1/2}G^{\mu \nu}\partial_\nu\right)$$
for a metric $G$.
If we consider the upper half-plane with metric
$$
G=\frac{1}{(1-\rho^2)f_2^2g_2^2}
\left[
\begin{array}{cc}
g_2^2&-\rho f_2 g_2\\
-\rho f_2 g_2&f_2^2
\end{array}
\right],\ \
G^{-1}=
\left[
\begin{array}{cc}
f_2^2&\rho f_2 g_2\\
\rho f_2 g_2&g_2^2
\end{array}
\right],\ \
g=\det{G}=
\frac{1}{(1-\rho^2)f_2^2g_2^2},
$$
then we have
$$\partial_\tau G=\left(\frac{1}{2}\Delta^\G+f\right)G,$$
where $f$ is a first-order operator. In order to apply Molchanov's
theorem\footnote{and more generally Minakshisundaram-Pleijel recursion formula.},
we need to be able to calculate the geodesics of the space with metric
$$
G=\frac{1}{(1-\rho^2)}
\left[
\begin{array}{cc}
\left(\frac{1}{f_2}\right)^2&-\rho \frac{1}{f_2} \frac{1}{g_2}\\
-\rho \frac{1}{f_2} \frac{1}{g_2}&\left(\frac{1}{g_2}\right)^2
\end{array}
\right].
$$

The purpose of this chapter is to understand the sufficient conditions for the geodesics to be easy to calculate.

\section{The isometry}

In order to be able to calculate the geodesics easily, we look for a suitable isometry. We impose the $\Sigma$-coordinate to be stable
(simpler case to keep the upper half-plane stable), so we look for an isometry on the form
$$\nabla\Phi=\left[
\begin{array}{cc}
\frac{1}{\sqrt{1-\rho^2}u}&\frac{-\rho}{\sqrt{1-\rho^2}v}\\
0&1
\end{array}
\right].$$
Then $H={(\nabla\Phi)^{-1}}^\top G{(\nabla\Phi)^{-1}}$ is diagonal iff
$$g_2 u=f_2 v.$$
We then have
$$
H=
\left[
\begin{array}{cc}
\frac{u^2}{f_2^2}&0\\
0&\frac{1}{g_2^2}
\end{array}
\right].
$$

For $\Phi$ to be well defined, the following condition is necessary :
$$\partial_\Sigma u=\partial_F v.$$

Interesting special cases are for $u=f_2$ and, more generally for  $u/f_2$  not dependant on $F$.

\section{Special cases}

In this part, we consider special cases showing the interest of this geometrical formulation. The first one confirms results about
non-stochastic volatility models. The second one gives a condition on $f_2$ and $g_2$ for both horizontal and vertical lines
to be geodesics, which is a very useful case. Finally, the third one shows that the $\delta$-model is the only \og $f_2$-separable\fg\ model for which
we have the homogeneity property.

\begin{itemize}
\item $g_2=0$ (no stochastic volatility) : then the geodesical distance between $(f_0,\sigma_0)$ and $(f,\sigma)$ is infinite, except if
$\sigma_0=\sigma$, so the probability of transition is concentrated on this line of the plane : $\sigma=\sigma_0$.
Varadhan's theorem in dimension one then gives
$$\log(P(f_0\to f))\underset{t\to 0}{\sim}\frac{-d^2}{2t}\mbox{ with }d=\left(\int_{f_0}^f\frac{\d x}{f_2(\sigma_0,x)}\right)^2.$$

\item we want to find an isometry between $\G$ and a simpler space $\H$. The most interesting is a function $\Phi$ such as
$$\nabla\Phi=\left[
\begin{array}{cc}
\frac{1}{\sqrt{1-\rho^2}f_2}&\frac{-\rho}{\sqrt{1-\rho^2}g_2}\\
0&1
\end{array}
\right],
$$
because we then have
$$G=(\nabla\Phi)^\top H (\nabla \Phi)\mbox{ with }H=
\left[
\begin{array}{cc}
1&0\\
0&\frac{1}{g_2^2}
\end{array}
\right].
$$
However, to find a convenient function $\Phi$, the partial derivatives must follow Schwarz's theorem of commutation (Poincaré's criterium
for the differential to be exact).
This is the case iff $f_2$ does not depend on $\Sigma$. So this case is simply not
pertinent\footnote{However, one can imagine new models where
$g_2$ depends on $F$, which seems quite natural. To be able to calculate the geodesics,
we then need the relation $\partial_\Sigma(1/f_2)=\partial_F(-\rho/g_2)$ ($*$),
which just necessitates well chosen functions. Moreover, the relation ($*$) gives interesting qualitative results : if the correlation is positive,
then the volvol decreases with $F$; the contrary if the correlation is negative.}.

\item if $u/f_2$ only depends on $\Sigma$, then with Schwarz's relation one can easily check that
the function $f_2$ is separable : let us write $f_2=a(F) b(\Sigma)$. Then
we also want to find an isometry between $\G$ and a simpler space $\H$. The good function $\Phi$ is such as
$$\nabla\Phi=\left[
\begin{array}{cc}
\frac{1}{\sqrt{1-\rho^2}a}&\frac{-\rho}{\sqrt{1-\rho^2}\frac{b}{g_2}}\\
0&1
\end{array}
\right],
$$
because we then have
$$G=(\nabla\Phi)^\top H (\nabla \Phi)\mbox{ with }H=
\left[
\begin{array}{cc}
\frac{1}{b^2}&0\\
0&\frac{1}{g_2^2}
\end{array}
\right].
$$
Schwarz's commutation relation is now true, so we really defined an isometry. We now just need to be able to calculate geodesics in the space $\H$.
First notice that the metric does not depend on the coordinate $x$, so, using Killing Fields, a system of equations for the geodesics is
$$
\left\{
\begin{array}{ccc}
\frac{\p{x}^2}{b(y)^2}+\frac{\p{y}^2}{g_2(y)^2}&=&c_1\\
\frac{\p{x}}{b(y)^2}&=&c_2\\
\end{array}
\right.,
$$
with $(x,y)=\Phi(F,\Sigma)$.
Thanks to this system, one can check that we have the homogeneity
property (if $\gamma$ is a geodesic, so is $\lambda \gamma$) iff
$b(y)=g_2(y)=y^\delta$ on the upper half-plane\footnote{
Indeed, we have such homogeneity for every geodesic iff the following system is satisfied for every
$\lambda$ and $y$ in $\R_*^+$ (we also suppose that the functions $b$ and $g_2$ are strictly positive and $\mathscr{C}^1$ on $\R_*^+$) :
$$
\left\{
\begin{array}{ccc}
\frac{b(\lambda y)}{g_2(\lambda y)}&=&\frac{b(y)}{g_2(y)}\\
b(\lambda y)&=&c(\lambda) b(y)
\end{array}
\right..
$$
This implies that we also have $g_2(\lambda y)=c(\lambda) g_2(y)$, so we only need to show that $b$ is a power.
Moreover, as $b(\lambda y)=c(\lambda) b(y)=c(y) b(\lambda)$, we have $b(y)/c(y)=cst$, so we only need to show
that $c$ is a power.

For $n\in\N^*$, we have $c(\lambda)^n b(y)=c(\lambda^{n-1})b(\lambda y)=\dots=c(\lambda)b(\lambda^{n-1}y)=b(\lambda^n y)=c(\lambda^n)b(y)$, so
$c(\lambda^n)=c(\lambda)^n$. The following is straightforward : we also have $c(\lambda^r)=c(\lambda)^r$ for $r\in\Q_+^*$, so, as $\Q_+^*$ is dense
in $\R_+$ and $c$ is continuous, we have $c(\lambda^s)=c(\lambda)^s$ for $s\in\R_+$. If $y\geq 1$, we then have
$c(y)=c(e^{\ln{y}})=c(e)^{\ln{y}}=y^{\ln{c(e)}}=y^{\delta_1}$.
For $0<y\leq 1$, we also have $c(y)=c((1/e)^{-\ln{y}})=c(1/e)^{-\ln{y}}=y^{-\ln{c(1/e)}}=y^{\delta_2}$. As $c$ is $\mathscr{C}^1$ at $y=1$, we
have $\delta_1=\delta_2$, so $c$, $b$ and $g_2$ can be written $y^\delta$.
}.
\end{itemize}

\section{Summary}

{\bf Proposition.} \it We consider all isometries of the upper half-plane which are identity for the $\sigma$-coordinate.
Then :
\begin{itemize}
\item if we want the resulting metric to be diagonal and only $\Sigma$-dependant,
then $f_2$ must be separable ($f_2=a(F)b(\Sigma)$). One then can effectively find a suitable isometry and the resulting metric is given by
$H=
\left[
\begin{array}{cc}
\frac{1}{b(\Sigma)^2}&0\\
0&\frac{1}{g_2(\Sigma)^2}
\end{array}
\right].$ If the geodesics follow the homogeneity property, then the only suitable metric is
$H=
\left[
\begin{array}{cc}
\frac{1}{\Sigma^{2\delta}}&0\\
0&\frac{1}{\Sigma^{2\delta}}
\end{array}
\right].$

\item one can find other isometries giving the following simpler metric : $H=
\left[
\begin{array}{cc}
1&0\\
0&\frac{1}{g_2^2}
\end{array}
\right].$
This necessitates the dependance $g_2(f,\sigma)$ (so this is not exactly a stochastic volatility model)
and the equation $\partial_\Sigma(1/f_2)=\partial_F(-\rho/g_2)$.

\end{itemize}

\rm

\chapter{The $\delta$-model : geometrical formulation}{}{}
\renewcommand{\theequation}{1.\arabic{equation}}

In order to have a geometrical vision of the situation, we first change the variables and the function (approximately like in Lesniewski's
preprint\footnote{{\sc Andrew Lesniewki,} {\it Probability Distribution in the SABR Model of Stochastic Volatility},
preprint}), writing :
\begin{itemize}
\item $s=\nu^2\tau$;
\item $x=f$;
\item $X=F$;
\item $y=\sigma/\nu$;
\item $Y=\Sigma/\nu$;
\item $K_{X,Y}(s,x,y)=\nu G_{X,\nu Y}(s/\nu^2,x,\nu y)$.
\end{itemize}
\vspace{0.3cm}

Now the Green function, solution of the PDE (2),
is solution of the following system (where we have $\lambda=\lambda'/\nu^2$ and $\mu=\mu'/\nu$):
\begin{equation}
\left\{
\begin{array}{ccl}
\frac{\partial K}{\partial s}&=&\frac{1}{2}\nu^{2\delta-2} y^{2\delta}\left[c(x)^2\frac{\partial^2 K}{\partial x^2}+2\rho c(x)\frac{\partial^2 K}{\partial x\partial y}+
\frac{\partial^2 K}{\partial x \partial y}\right]+\lambda(y-\mu)\frac{\partial K}{\partial y}\\
K_{X,Y}(0,x,y)&=&\delta(x,X)\ \delta(y,Y)
\end{array}
\right..
\end{equation}

In this part, we first rewrite the PDE (1.1) in terms of the Laplace
Beltrami operator of a suitable space (the important properties of this operator are summarized in appendix 2).
We then use an isometry that
allows us to consider a simpler generic space. Our results are
summarized in a final table.\\

\boldmath
\section{Suitable form for the partial differential equation}
\unboldmath

The metric associated with the PDE (1.1) is
$$G=\frac{1}{\nu^{2\delta-2}(1-\rho^2)y^{2\delta}c(x)^2}
\left[
\begin{array}{cc}
1&-\rho c(x)\\
-\rho c(x)&c(x)^2
\end{array}
\right].
$$

We call the
corresponding Riemannian space $\G$. We also need the determinant $g$ of $G$ and the inverse matrix :
$$g=\frac{1}{\nu^{4\delta-4}(1-\rho^2)y^{4\delta}c(x)^2},\ \
G^{-1}= \nu^{2\delta-2}y^{2\delta}
\left[
\begin{array}{cc}
c(x)^2&\rho c(x)\\
\rho c(x)&1
\end{array}
\right].
$$

The Laplace Beltrami operator is defined as
$$\Delta^\G=g^{-1/2}\partial_\mu\left(g^{1/2}g^{\mu \nu}\partial_\nu\right).$$
A simple calculation gives $\Delta^\G=\nu^{2\delta-2}y^{2\delta}(c(x)^2\partial_{x x}+2\rho c(x) \partial_{x y}+\partial_{y y})
+\nu^{2\delta-2}y^{2\delta}c(x)c'(x)\partial_x$. This is the reason why (1.1), with the suitable normalized initial condition,
can be written as
\begin{equation}
\left\{
\begin{array}{ccc}
\frac{\partial K^{\G}}{\partial s}&=&\left(\frac{1}{2}\Delta^\G + f^\G\right)K^\G\\
K^\G_Z(0,z)&=&\delta(z,Z)
\end{array}
\right.,
\end{equation}
with
\begin{equation}
f^\G=-\frac{1}{2}\nu^{2\delta-2}y^{2\delta}c(x)c'(x)\partial_x+\lambda(y-\mu)\partial_y.
\end{equation}
and $\delta(z,Z)=\delta(x,X)\ \delta(y,Y)$.
The interest of this formulation is that we have separated the \og intrinsic\fg~Laplacian from the first-order term (without
changing variables like in Hagan's paper\footnote{{\sc Patrick S. Hagan, Deep Kumar, Andrew S. Lesniewski and Diana E. Woodward}, {\it Managing Smile Risk}.}).

\boldmath
\section{The isometry}
\unboldmath

We follow here the general methodology of the chapter 1.
Let us introduce $\H$, the $\delta$-space  with metric
$$
H=\left[
\begin{array}{cc}
\frac{1}{y^{2\delta}}&0\\
0&\frac{1}{y^{2\delta}}
\end{array}
\right].$$

Let us $\Phi$ be the application of the upper half-plane
$$\Phi\left(
\left[
\begin{array}{c}
x\\
y
\end{array}
\right]
\right)=
\frac{1}{\nu^{\delta-1}}\left[
\begin{array}{c}
\frac{1}{\sqrt{1-\rho^2}}\left(\int_p^x\frac{\d u}{c(u)}-\rho y\right)\\
y
\end{array}
\right]
,$$
where $p$ is a positive constant. Then the jacobian matrix of $\Phi$ is
$$
\nabla\Phi=
\frac{1}{\nu^{\delta-1}}
\left[
\begin{array}{cc}
\frac{1}{\sqrt{1-\rho^2}c(x)}&-\frac{\rho}{\sqrt{1-\rho^2}}\\
0&1
\end{array}
\right],
$$
so one can easily check that between the metrics the following relationship holds :
$$G=(\nabla\Phi)^\top
H
\nabla\Phi.
$$

This relationship exactly means that $\Phi$ conserves the scalar product : in $\G$, for two trajectories $\gamma^1$ and $\gamma^2$ crossing in $z$ at $t$,
$<\d/\d t\ \Phi(\gamma_1^t),\d/\d t\ \Phi(\gamma_2^t)>_\H=(\nabla\Phi(z)\p{\gamma_1^t})^\top H(\nabla\Phi(z)\p{\gamma_2^t})=
{\p{\gamma_1^t}}^\top G {\p{\gamma_2^t}}=
<\d/\d t\ \gamma_1^t,\d/\d t\ \gamma_2^t>_\G$.
This local property implies the global result that $\Phi$ induces an isometry between $\G$ and $\H$.\\

The following paragraphs are quite technical : in order to simplify our problem, we find a  diffusion equation
in $\H$ equivalent to the one in $\G$. We then find the relations between the transition probabilities associated to these diffusion equations.

\boldmath
\section{Reformulation of the hyperbolic problem}
\unboldmath

We now consider the following Green problem in $\H$ :
\begin{equation}
\left\{
\begin{array}{ccccc}
\frac{\partial K^\H}{\partial s}&=&\left(\frac{1}{2}\Delta^\H+f^\H\right) K^\H&&\\
K^\H_{\Phi(Z)}(0,z')&=&\delta(z',\Phi(Z))\ &=& \det\left[\nabla\Phi(Z)\right]^{-1}\delta(z,Z)
\end{array}
\right.,
\end{equation}
with :
\begin{itemize}
\item $z'=\Phi(z)$;
\item $\Delta^\H$ the Laplace Beltrami operator corresponding to the Riemannian space $\H$, therefore $\Delta^\H=y^{2\delta}(\partial_{xx}+\partial_{yy})$;
\item $f^\H(A)=\nabla\Phi(\Phi^{-1}(A))f^\G(\Phi^{-1}(A))$.
\end{itemize}

If $K^\H$ is a solution of (1.4), then
$\det\left[\nabla\Phi(Z)\right] K^\H\circ\Phi$ is a solution of (1.3) ($\det\left[\nabla\Phi(Z)\right]$, which is a normalization
constant, is written $C$ in the following). Indeed, as $\Phi$ does not modify the time $s$,
\begin{eqnarray*}
\frac{\partial K^\G}{\partial s}&=&\frac{\partial (C\ K^\H\circ\Phi)}{\partial s}\\
&=&C \left(\frac{\partial K^\H}{\partial s}\right)\circ\Phi\\
&=&C\left(\left(\frac{1}{2}\Delta^\H+f^\H\right) K^\H\right)\circ \Phi\\
&=&C\ \left(\frac{1}{2}\Delta^\H K^\H\right)\circ\Phi+C\ \nabla \Phi f^\G(K^\H)\\
&=&C\ \frac{1}{2}\Delta^\G(K^\H\circ \Phi)+C\ f^\G(K^\H\circ\Phi)\\
\frac{\partial K^\G}{\partial s}&=&\left(\frac{1}{2}\Delta^\G +f^\G\right)K^\G.
\end{eqnarray*}

Thanks to the normalization constant $\det\left[\nabla\Phi(Z)\right]$, $K^\G$ follows the condition at $s=0$, so
from the uniqueness of the solution of
(1.3) we deduce that
\begin{equation}
K^{\G}=\det\left[\nabla\Phi(Z)\right]K^\H\circ\Phi.
\end{equation}

\boldmath
\section{From density to probability of transition}
\unboldmath

The quantity $K^\H$ is a density with respect to the euclidean metric.
To be consistant with the next chapter, we now note $K$ the corresponding density probability with respect to the metric $H$ :
$$K=\frac{K^\H}{\sqrt{\det{H}}}=y^{2\delta}K^\H.$$

We have then
\begin{eqnarray*}
P&=&\int_{\R }G\d \Sigma\\
&=&\int_{\R }K^\G(X^\G,Y^\G)\d Y^\G\\
&=&\int_{\R }K^\H\circ\Phi(X^\G,Y^\G)\ \det{\nabla\Phi(X^\G,Y^\G)}\ \d Y^\G\\
&=&\int_{\R }\frac{K\circ\Phi(X^\G,Y^\G)}{{Y^\H}^{2\delta}}\ \det{\nabla\Phi(X^\G,Y^\G)}\ \d Y^\G\\
&=&\int_{\R }\frac{K\circ\Phi(X^\G,Y^\G)}{{Y^\H}^{2\delta}} \frac{1}{\nu^{2\delta-2}\sqrt{1-\rho^2}c(X^\G)} \nu^{\delta-1}\d Y^\H\\
P&=&\frac{1}{\nu^{\delta-1}\sqrt{1-\rho^2}c(X^\G)}\int_{\mathbb{D}}\frac{K(X^\H,Y^\H)}{{Y^\H}^{2\delta}}\ \d Y^\H
\end{eqnarray*}
where $\mathbb{D}$ is the image of the straight line $x=X^\G$ by the application $\Phi$, that is to say the straight line with equation
$$x=\frac{1}{\nu^{\delta-1}\sqrt{1-\rho^2}}\left(\int_{p}^{X^\G} \frac{\d u}{c(u)}-\rho \nu^{\delta-1}y\right).$$
So if we define
\begin{equation}
P^\H=\int_{\mathbb{D}}\frac{K(X^\H,Y^\H)}{{Y^\H}^{2\delta}}\d Y^\H,
\end{equation}
then
$$P=\frac{1}{\nu^{\delta-1}\sqrt{1-\rho^2}c(X^\G)}\ P^\H.$$

In the following we will need the expression of the moment of order $2\delta$ of $G$, that is
to say
\begin{equation}
M=\int_{\R }G \Sigma^{2\delta}\d \Sigma
\end{equation}
If we define
\begin{equation}
M^\H=\int_{\mathbb{D}}K(X^\H,Y^\H)\d Y^\H,
\end{equation}
then the same calculation as above gives
$$
M=\frac{\nu^{2\delta^2}}{\nu^{\delta-1}\sqrt{1-\rho^2}c(X^\G)}M^\H.
$$

Here is a table that summarizes the results of the first part.

\scriptsize
\begin{center}
\begin{tabular}{lcc}
\hline\\[-1mm]
\multicolumn{3}{c}{\bfseries From $\G$ to $\H$}\\\\[-2mm]
\hline\\[-2mm]
The space &  $\G$ & $\H$\\\\[-2mm]
The metric &$G=\frac{1}{\nu^{2\delta-2}(1-\rho^2)y^{2\delta}c(x)^2}
\left[
\begin{array}{cc}
1&-\rho c(x)\\
-\rho c(x)&c(x)^2
\end{array}
\right]$&$$
$H=\left[
\begin{array}{cc}
\frac{1}{y^{2\delta}}&0\\
0&\frac{1}{y^{2\delta}}
\end{array}
\right]$\\\\[-2mm]
\hline\\[-2mm]
The isometry&\multicolumn{2}{c}{$\Phi\left(
\left[
\begin{array}{c}
x\\
y
\end{array}
\right]
\right)=
\frac{1}{\nu^{\delta-1}}\left[
\begin{array}{c}
\frac{1}{\sqrt{1-\rho^2}}\left(\int_p^x\frac{\d u}{c(u)}-\rho y\right)\\
y
\end{array}
\right]
$}\\\\[-2mm]
The LBO&$\begin{array}{lll}\Delta^\G&=&\nu^{2\delta-2}y^{2\delta}(c(x)^2\partial_{x x}\\&+&2\rho c(x) \partial_{x y}+\partial_{y y})\\
&+&\nu^{2\delta-2}y^{2\delta}c(x)c'(x)\partial_x\end{array}$&$\Delta^\H=y^{2\delta}(\partial_{xx}+\partial_{yy})$\\\\[0mm]
The first-order operator&$f^\G=
\left[\begin{array}{c}
-\frac{1}{2}\nu^{2\delta-2}y^{2\delta}c(x)c'(x)\\ \lambda(y-\mu)
\end{array}
\right]$
&$f^\H(A)=\nabla\Phi(\Phi^{-1}(A))f^\G(\Phi^{-1}(A))$\\\\[-2mm]
\hline\\[-2mm]
The diffusion equations&$\left\{
\begin{array}{ccc}
\frac{\partial K^{\G}}{\partial s}&=&\left(\frac{1}{2}\Delta^\G + f^\G\right)K^\G\\
K^\G_Z(0,z)&=&\delta(z,Z)
\end{array}
\right.$&$\left\{
\begin{array}{ccc}
\frac{\partial K^\H}{\partial s}&=&\left(\frac{1}{2}\Delta^\H+f^\H\right) K^\H\\
K^\H_{\Phi(Z)}(0,z')&=&\delta(z',\Phi(Z))
\end{array}
\right.$\\\\[-2mm]
Link between $K^\G$ and $K^\H$&\multicolumn{2}{c}{$K^{\G}=\det\left[\nabla\Phi(Z)\right]K^\H\circ\Phi$}\\\\[-2mm]
\hline\\[-2mm]
The transition probability&$P=\frac{1}{\nu^{\delta-1}\sqrt{1-\rho^2}c(X^\G)}\ P^\H$&$P^\H=\int_{\mathbb{D}}\frac{K(X^\H,Y^\H)}{{Y^\H}^{2\delta}}\d Y^\H$ \\\\[-2mm]
\hline\\[-2mm]
The moment &$M=\frac{\nu^{2\delta^2}}{\nu^{\delta-1}\sqrt{1-\rho^2}c(X^\G)}\ M^\H$&$M^\H=\int_{\mathbb{D}}K(X^\H,Y^\H)\d Y^\H$ \\\\[-2mm]
\hline
\end{tabular}
\end{center}
\normalsize

All relationships above imply that we just need to get a suitable approximation for $K^\H$. This is the purpose of the next chapter.

\chapter{From Molchanov's theorem to the $\delta$-model}{}{}
\setcounter{equation}{0}
\renewcommand{\theequation}{2.\arabic{equation}}

We have shown in the previous part that it is sufficient to calculate all probability densities in $\H$, the $\delta$-space. This can be done thanks
to the following theorem, expressed in terms of geometric invariants. The necessary calculations and adaptations to the $\delta-$space will be done after.

\boldmath
\section{Molchanov's theorem : the 0-order expansion}
\unboldmath

Molchanov's theorem gives an asymptotic behavior for the density transition function, for small maturities. More precisely, we use the following notations :

\begin{itemize}
\item $l$ is the dimension of the space;
\item $H$ is a specified metric of a space $\H$;
\item $\Delta^\H$ is the Laplace-Beltrami operator corresponding to the metric $H$;
\item $f$ is a first-order operator;
\item $K(s,a,b)$ is the density function (with respect to the element of volume $v^\H$) corresponding to the diffusion process
$\frac{\partial K}{\partial s}=\left(\frac{1}{2}\Delta^\H + f\right)K$
with initial condition $K(0,a,b)=\delta(a,b)$ :
$K(s,a,b)$ represents the probability density of getting in $b$, leaving from $a$, after a time $s$, for the metric $H$.
\end{itemize}

Supposing sufficient conditions of regularity (uniqueness of the geodesic between two points\dots), we can now formulate the theorem we need.\\

\fbox{
\begin{minipage}{0.9\textwidth}
\bf{Molchanov's theorem.} \it Under all the conditions above, if $d(a,b)$ is the distance between $a$ and $b$ along the unique geodesic linking them, $\gamma$, then
$$K(s,a,b)\underset{s\to 0}{\sim}\frac{1}{(2\pi s)^{l/2}}e^{-\frac{d(a,b)^2}{2s}+W(a,b)}\sqrt{\frac{d(a,b)^{l-1}}{\Psi(a,b)}}$$
with
\begin{itemize}
\item $\Psi(a,b)=\d S/\d \varphi$ where a cone of light from $x$ with a solid angle $\d \varphi$ illuminates an area $\d S$ on the geodesic hypersurface
orthogonal to $\gamma$ at $y$ (cf the following figure);
\item $W(a,b)=\int_{0}^{d(a,b)}<f, \d\gamma/\d s>\d s$ : this is the work of the field $f$ along the geodesic.
\end{itemize}
\end{minipage}
}
\vspace{0.3cm}
\rm
\begin{proof}
A complete proof can be found in Molchanov's article\footnote{{\sc S. A. Molchanov}, {\it Diffusion Processes and Riemannian Geometry}, Russian Math. Surveys 30 : 1 (1975).}.
However, one can give an intuition of this result if there is no first-order operator, by analogy with the euclidean case.

\begin{wrapfigure}[13]{r}{.5\linewidth}
\psset{unit=0.6cm}
\begin{center}
\begin{pspicture}(0,1)(11,6.5)
\pscurve{->}(4,3)(7,2)(9,1)
\pscurve{->}(4,3)(7,4.75)(8,6)
\pscurve{->}(4,3)(5,4.25)(6.5,6.3)
\pscurve{->}(4,3)(4.2,5)(4.5,6.5)
\pscurve{->}(4,3)(3.2,5)(3,6)
\pscurve{->}(4,3)(3,3.5)(1.5,4.5)
\pscurve{->}(4,3)(3,2.8)(1.5,2.5)
\pscurve{->}(4,3)(3,2.2)(2.5,1.5)
\pscurve{->}(4,3)(3.9,2)(4,0.7)
\pscurve{->}(4,3)(5,1.75)(7,0.2)
\psccurve[linestyle=dashed](9,1)(8,6)(6.5,6.3)(4.5,6.5)(3,6)(1.5,4.5)(1.5,2.5)(2.5,1.5)(4,0.7)(7,0.2)
\rput(4,3){\psframebox*[framearc=.3]{$O$}}
\psarc(4,3){1}{-18}{28}
\pscurve{<->}(9.3,1)(9.6,2)(9.4,4.2)(8.3,6.2)
\rput(10.2,3.4){$\d S$}
\rput(6.2,3){$\d \varphi$}
\rput(8.8,3.5){$K$}
\rput(9.3,3.5){\textbullet}
\end{pspicture}
\end{center}
\caption{Geodesic surface for a curved space}
\end{wrapfigure}
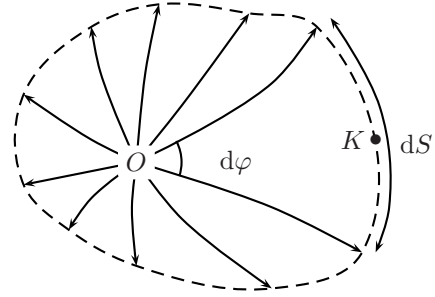

For a flat space, the above formula is clearly true, and for a curved space the density naturally decreases with
$\d S /\d \varphi$, where $\d \varphi$ is the initial angle between geodesics leaving $O$. Indeed, imagine a diffusion in this space, materialized with particles.
The particles are diffused in an isotropic way
from $O$, and they remain blocked between the geodesics, so $K \d S$ is analogous to a one-dimension diffusion
density with a density $K_0 \d \varphi$ in $O$.

 This does not leads us to the exact formula (because particles come from the other side of $\d \varphi$), but it
gives an intuition of the origin of the function $\Psi$.\\

From a computational point of view, we need a definition of $\Psi$ relatively to coordinates.
Let $(e_i)$ be a base of the orthogonal to $\gamma$ in $O$. Let us $\mathscr{J}_i(t)$ ($1\leq i\leq l-1$,
$1\leq t\leq d(a,b)$) be the
Jacobi field\footnote{see {\sc Jûrgen Jost,} {\it Riemannian
Geometry and Geometric Analysis,} Springer Universitext, 2002.} along $\gamma$ such as $\mathscr{J}_i(0)=0$ and $\mathscr{J}_i'(0)=e_i$, and $\mathscr{Z}(t)=
\left[\mathscr{J}_1(t),\dots,\mathscr{J}_{l-1}(t)\right]$. Then
$$\Psi(x,y)=\det{\mathscr{Z}(d(a,b))}.$$

Indeed, $\det\left[f_1,\dots,f_{l-1}\right]$ represents the area engendered by a polyhedra of sides $f_1,\dots,f_{l-1}$,
so $\det\mathscr{Z}(t)$ represents the infinitesimal area engendered by the geodesics at time $t$, fo a solid angle $\d \varphi$.
\end{proof}
\vspace{0.3cm}

What we will need in reality is both of the following corollaries.\\

\fbox{
\begin{minipage}{0.9\textwidth}
{\bf\boldmath Corollary 1 : dimension 2 (for the $\delta$-model).\unboldmath} \it In the plane, we consider the following system,
$$
\left\{
\begin{array}{ccc}
\overset{..}{Z}+R(t) Z&=&0\\
Z(0)&=&0\\
\overset{.}{Z}(0)&=&1
\end{array}
\right.,
$$
where $R(t)$ is the Gauss curvature for the point $\gamma(t)$ of the geodesic linking $a$ and $b$. Then
$$K(s,a,b)\underset{s\to 0}{\sim}\frac{1}{2\pi s}e^{-\frac{d(a,b)^2}{2s}+W(a,b)}\sqrt{\frac{d(a,b)}{Z(d(a,b))}}.$$
\end{minipage}
}
\rm

\vspace{0.3cm}
\begin{proof}
We only need to show that the the evolution of the length $Z$ of the unique vector of $\mathscr{Z}$ is
governed by the equation above. The evolution equation for a Jacobi field is
$$\pp{y^k} +y^j(t)R[e_j(t),\p{\gamma},\p{\gamma},e_k(t)]=0$$
for each of the $l-1$ composants $y_k$,
where :
\begin{itemize}
\item $R$ is the Riemann tensor;
\item $\left\{e_1(0),\dots,e_{l-1}(0)\right\}$ is a base of the orthogonal hyperplane of $\gamma$ at $a$;
\item $e_i(t)$ is the parallel transport of $e_i(0)$ along $\gamma$ relatively to the field $\p{\gamma}$;
\item $y_i$ is the composant of the Jacobi field on $e_i$.
\end{itemize}
\vspace{0.3cm}

We also know that the parallel transport relatively to $\p{\gamma}$ transforms $e_1(0)$ in
the orthogonal of $\p{\gamma}$, with norm $|\p{\gamma}|$, written $\p{\gamma^\perp}$. So the evolution equation can be written,
in dimension 2,
$$\pp{y} +y R[\p{\gamma^\perp},\p{\gamma},\p{\gamma},\p{\gamma^\perp}]=0.$$

From the definition of the Gauss curvature, $R[\p{\gamma^\perp},\p{\gamma},\p{\gamma},\p{\gamma^\perp}]=R(t)$,
which completes the proof.
\end{proof}

\vspace{0.3cm}
\fbox{
\begin{minipage}{0.9\textwidth}
\bf{\boldmath Corollary 2 : dimension 2 and constant Gauss curvature (for the $\delta$-model with $\delta=1$ : SABR).\unboldmath} \it If the plane has a constant gaussian curvature $-k<0$, then
$$K(s,a,b)\underset{s\to 0}{\sim}\frac{1}{2\pi s}e^{-\frac{d(a,b)^2}{2s}+W(a,b)}
\sqrt{\frac{\sqrt{k}\ d(a,b)}{\sinh{\sqrt{k} d(a,b)}}}.$$
\end{minipage}
}
\rm
\vspace{0.3cm}

\begin{proof}
Immediate from the corollary 1.
\end{proof}

\boldmath
\section{The first-order expansion}
\unboldmath

We can get a better approximation for the probability density $K$. Indeed, one can show that for a sufficiently regular diffusion and space,
$K$ admits the following expansion\footnote{see {\sc Elton P. Hsu,}
{\it Stochastic Analysis on Manifolds,} American Mathematical Society vol. 38,
Graduate Studies in Mathematics, 2001, theorem 5.1.1 p.130.}, for any $n\in\mathbb{N}$ :
$$K(s,a,b)=\frac{1}{(2\pi s)^{l/2}}e^{-\frac{d(a,b)^2}{2s}+W(a,b)}\sqrt{\frac{d(a,b)^{l-1}}{\Psi(a,b)}}(1+K_1(a,b)s+\dots+K_n(a,b)s^n+O(s^{n+1})).$$

Our purpose is to give the equation verified by $K_1$ in the case of the $\delta-$geometry, and then evaluate its solution for a point $b$ near from $a$.


The expansion above can be written at first order as
\begin{equation}
K=\frac{1}{s}e^{-\frac{d^2}{2s}}g(1+K_1 s +O(s^2)).
\end{equation}

Injecting (2.1) in the diffusion equation $\partial_s K=\left(1/2\ \Delta^\H + f\right)K$, we successively get the
following diffusion equations satisfied by the functions
$d$, $g$ and $K_1$, for the diffusion in the $\delta-$space :
\begin{eqnarray}
y^{2\delta}|\nabla d|^2&=&1,\\
g&=&2 d\ g \left(\frac{1}{2}\Delta^\H+f\right)d+y^{2\delta}\nabla g\cdot\nabla d^2,\\
g\ K_1&=& \left(\frac{1}{2}\Delta^\H+f\right)g +y^{2\delta}d\ g\ \nabla K_1 \cdot\nabla g.
\end{eqnarray}

Here $|\ |$, $\cdot$ and $\nabla$ reference respectively the euclidean norm, the euclidean scalar product and the euclidean gradient ($\nabla=(\partial_x,\partial_y)$), and
$\Delta^\H=y^{2\delta}(\partial_{xx}+\partial_{yy})$. These equations need some comments :
\begin{itemize}
\item (2.2) just traduces that $\d s^2=(\d x^2+\d y^2)/y^{2\delta}$;
\item (2.3) is a transport equation that can be solved by integration along the geodesic : the explicit solution is given by Molchanov's theorem :
$$g(a,b)=\frac{1}{2\pi}e^{W(a,b)}\sqrt{\frac{d(a,b)}{Z(d(a,b))}}.$$
\item  (2.4) is also a transport equation, that can be solved by integration along lines directed by $\nabla g$ instead of $\nabla d$ : the knowledge of the geodesics is not sufficient anymore.
This is the reason why we will just calculate its initial value ($K_1(a,a)=(K_1)_0$), which will be a sufficient correction for us in the following.
\end{itemize}

A great calculation (see appendix 2.3) gives us the following result for $(K_1)_0$ :
$$(K_1)_0=\frac{-\delta y_0^{2\delta-2}}{6}+\frac{(\operatorname{div} f)_0}{2}-\frac{(f_y)_0}{y_0}+\frac{3}{2}\frac{|f_0|^2}{y_0^{2\delta}}.$$

The main result of this part is then the following expansion, in the $\delta$-space :

$$\boxed{
K(s,a,b)=\frac{1}{2\pi s} e^{-\frac{d(a,b)^2}{2s}+W(a,b)}\sqrt{\frac{d(a,b)}{Z(d(a,b))}}\left(1+(K_1)_0 s+O(s^2)\right)}$$
with :
\begin{itemize}
\item the functions $d$ and $Z$ computed in $O(1)$, thanks to the homogeneity property of the $\delta$-space (cf. appendix 2.4);
\item the function $W$ easily computed as the integral of a known function along a known path.
\item $(K_1)_0$ given by the formula above.
\end{itemize}

{\bf Remark 1.} When $f=0$, our result for $(K_1)_0$ is a special case of a general formula :
in any riemannian space we have
$$(K_1)_0=\frac{R}{6},$$
where $R$ is the scalar curvature.\\

{\bf Remark 2.} A formula by Minakshisundaram and
Pleijel\footnote{{\sc S. Minakshisundaram and Pleijel},
{\it Some properties of the eigenfunctions of the Laplace-operator on Riemannian manifolds},
Canadian J. Math. 1 (1949), 242-256.}
gives a general asymptotic expansion for $K$ to any order,
but it is too complex and non effective for the second order.


\chapter{Applications of the previous asymptotics}{}{}
\setcounter{equation}{0}

In this chapter, we integrate the asymptotic expression obtained previously.
This leads directly to digital option prices.

Then, using Hagan's local volatility formula\footnote{{\sc Patrick S. Hagan and Diana E. Woodward,} {\it Equivalent Black Volatilities},
Applied Mathematical Finance 6, 147-157 (1999).}, we compute, through equivalent local volatility, vanilla option prices.

\boldmath
\section{The Laplace transformation}
\unboldmath

We want to evaluate $P^\H$ and $M^\H$ defined by (1.6) and (1.8). We
use here a Laplace expansion of order 2 : for $f$ and $g$
sufficiently regular functions, and $d$ with minimum at $u_0$ and
non-zero second derivative, we have
\begin{eqnarray*}
\int_\R e^{-\frac{\Phi(u)}{s}}(f(u)+s\ g(u))\d u &=&
\sqrt{\frac{2\pi
s}{\Phi''(u_0)}}\Biggl(f(u_0)+s\Biggl(g(u_0)+\frac{f''(u_0)}{2\Phi''(u_0)}
-\frac{\Phi^{(4)}(u_0)f(u_0)}{8\Phi''(u_0)^2}\\
&-&\frac{f'(u_0)\Phi^{(3)}(u_0)}{2\Phi''(u_0)^2}+\frac{5\Phi^{(3)}(u_0)^2f(u_0)}{24\Phi''(u_0)^3}\Biggl)+O(s^2)\Biggl).
\end{eqnarray*}

In the case of $P^\H$ and $M^\H$ we need to calculate the ordinate $Y_{min}$ of $\mathbb{D}$ such as
$d(Z_0,Z_{min})$ is minimal. This is done in appendix 2.5.

\begin{wrapfigure}[10]{r}{.6\linewidth}
\psset{unit=0.4cm}
\begin{center}
\begin{pspicture}(-10,0)(10,8)
\psplot{-10}{10}{2 x 4 div 2 exp neg exp 10 mul}
\psplot[linestyle=dashed]{-10}{10}{2 x 4.3 div 2 exp neg exp 9 mul}
\psline{->}(-2,0)(10,0)
\psline{->}(-2,0)(-2,5)
\psline[linestyle=dotted](2,0)(2,8.5)
\rput(2,-0.7){$K$}
\rput(-2,6){$P$}

\end{pspicture}
\end{center}
\end{wrapfigure}

Unfortunately, we are unable to calculate the third and fourth variations of $\Phi$ for
any strike, but this can be done for $K=f_0$, at the money;
including these results in the 1-order term above will be sufficient : the effect of the 1-order correction is more
important a the money.

 To sum up, we will take the following approximation for $P$ :
\begin{multline*}
P \approx
\sqrt{\frac{2\pi
s}{\Phi''(u_0)|^{K}}}\Biggl(f(u_0)|^{K}+s\Biggl((K_1)_0+\frac{f''(u_0)|^{K=f_0}}{2\Phi''(u_0)|^{K=f_0}}
-\frac{\Phi^{(4)}(u_0)|^{K=f_0}}{8\Phi''(u_0)^2|^{K=f_0}}\\
-\frac{f'(u_0)|^{K=f_0}\Phi^{(3)}(u_0)|^{K=f_0}}{2\Phi''(u_0)^2|^{K=f_0}}+\frac{5\Phi^{(3)}(u_0)^2|^{K=f_0}}{24\Phi''(u_0)^3|^{K=f_0}}\Biggl)\Biggl).
\end{multline*}
with the following formulas :
\newpage
\begin{itemize}
\item $\frac{\d^2}{\d y^2}d(Y_{min})=
\frac{1}{\sin^2\theta_1}\left(\frac{\p{Z}(d)}{Z(d)Y_{min}^{2\delta}}\pm
\frac{\delta}{Y_{min}^{\delta+1}}\cos\theta_1\right),$
where the sign depends on the following configurations.
\psset{unit=0.5cm}
\begin{center}
\begin{pspicture}(0,0)(10,5)
\psline[linestyle=dashed](5,0)(5,3.5)
\psline(1,1)(4,4)
\psline[linestyle=dashed](1,1)(3,1)
\psarc(1,1){1}{0}{45}
\rput(2.6,1.5){$\theta_1$}
\psline(6,1)(9,4)
\pscurve(0.5,3.5)(1.5,3.7)(2.5,3.4)(3,3)
\pscurve(8,3)(8.5,2.4)(9,0.5)
\rput(2.5,4.5){$-$}
\rput(7.5,4.5){$+$}
\angleplat{5}{5}{3}{3}{2}{4}{0.4}
\angleplat{10}{5}{8}{3}{9}{2}{0.4}
\end{pspicture}
\end{center}

This second variation for the $\delta$-model is the most important result of this part because it gives
the leading order for pricing the digital options.
\item $\Phi''=d d''\underset{d\to 0}{\to}\frac{1}{\sin^2{\theta_1}Y_{0}^{2\delta}}$;
\item $\Phi^{(3)}|^{K=f_0}=\frac{-3\delta}{{Y_0}^{2\delta+1}\sin^2{\theta_1}}$;
\item $\Phi^{(4)}|^{K=f_0}=\frac{\delta(4+7\delta)}{Y_0^{2\delta+2}\sin^2{\theta_1}}.$
\end{itemize}
These results are justified in appendix 2.5.

\boldmath
\section{The transition probability : pricing digital options}
\unboldmath

All the previous results give a methodology to calculate the transition probability $P$ for any $\delta$. For instance, in the case
$\delta=1$, the explicit calculations are done at the end of this paragraph. What is done here for $\delta=1$ can be done for any $\delta$ : the trajectories of
$\delta$-geometry and the Jacobi fields calculations are given in appendix.

\boldmath
\subsection{Calculation of the field $f^\H$ and its work}
\unboldmath

If $c(x)=x^{\beta}$ ($0\leq \beta<1$ : the case $\beta=1$, needed for Heston's model, can be handled apart, the same way),
the function $\Phi$ has the following inverse,
$$\Phi^{-1}
\left(
\left[
\begin{array}{c}
X^\H\\
Y^\H
\end{array}
\right]
\right)=
\left[
\begin{array}{c}
\left((1-\beta)\left(\sqrt{1-\rho^2}X^\H+\rho Y^\H\right)\right)^\frac{1}{1-\beta}\\
Y^\H
\end{array}
\right]
.$$

Moreover,
\begin{eqnarray*}
\nabla\Phi(\gamma^\G(t))f(\gamma^\G(t))&=&
\left[
\begin{array}{cc}
\frac{1}{\sqrt{1-\rho^2}c(X^\G(t))}&-\frac{\rho}{\sqrt{1-\rho^2}}\\
0&1
\end{array}
\right]
\left[
\begin{array}{c}
-\frac{1}{2} {Y^\G}^{2\delta}c(X^\G)c'(X^\G)\\
\lambda(Y^\G-\mu)
\end{array}
\right]\\
&=&
\left[
\begin{array}{c}
\frac{1}{\sqrt{1-\rho^2}}\left(-\frac{1}{2}{Y^\G}^{2\delta}\beta {X^\G}^{\beta-1}-\rho\lambda(Y^\G-\mu)\right)\\
\lambda(Y^\G-\mu)
\end{array}
\right]\\
f^\H(\gamma^\H(t))&=&
\left[
\begin{array}{c}
\frac{1}{\sqrt{1-\rho^2}}\left(\frac{-(Y^\H)^{2\delta}\beta}{2(1-\beta)\left(\sqrt{1-\rho^2}X^\H+\rho Y^\H\right)}-\rho\lambda\left(Y^\H-\mu\right)\right)\\
\lambda\left(Y^\H-\mu\right)
\end{array}
\right].
\end{eqnarray*}

From $\delta$-geometry, we know that the hyperbolic space (ie the case $\delta=1$) has constant Gauss curvature : this specificity
makes all its interest, we can apply the corollary 2.
We know that geodesics are circles, so
$Y^\H(X)=\sqrt{({X^\H}_0-{X^\H}_{fin})^2+{{Y^\H}_0}^2-(X^\H-{X_{fin}}^\H)^2}$.
We can then calculate explicitly the work $W$ of $f^\H$ on these circles.
Simplifying by $\d t$, we get
\begin{eqnarray*}
W&=&\int_{\gamma^{\H}}
\frac{1}{\sqrt{1-\rho^2}}\left(\frac{-(Y^\H)^{2\delta}\beta}{2(1-\beta)\left(\sqrt{1-\rho^2}X^\H+\rho Y^\H\right)}-\rho\lambda\left(Y^\H-\mu\right)\right)\d X^{\H}
+
\int_{\gamma^{\H}}
\lambda\left(Y^\H-\mu\right)\d Y^{\H}\\
&=&\frac{-\beta}{2\sqrt{1-\rho^2}(1-\beta)}\int_{\gamma^{\H}}\frac{(Y^\H)^{2\delta}}{\sqrt{1-\rho^2}X^\H+\rho Y^\H}\d X^{\H}
-
\frac{\rho \lambda}{\sqrt{1-\rho^2}}\int_{\gamma^{\H}}Y^\H\d X^\H\\&+&\frac{\rho \lambda\mu}{\sqrt{1-\rho^2}}(X_{end}-X_0)+
\frac{\lambda}{2}(Y_{end}^2-Y_0^2)-\lambda\mu(Y_{end}-Y_0).
\end{eqnarray*}
This expression does not depend of the choice of time, which is normal for the work of a field on a trajectory.

\boldmath
\subsection{The 1-order expression of $P$ : the SABR case}
\unboldmath

We first need a primitive (in $x$) of the function
$$g(a,\rho,x)=\frac{1-x^2}{a+\sqrt{1-\rho^2}x+\rho\sqrt{1-x^2}}.$$
We will note it as $G(a,\rho,x)$. An explicit formula for $G$ is
\begin{eqnarray*}
G(a,\rho,x)&=&
-\frac{1}{4\sqrt{a^2-1}}\Biggl(
-8 a \rho(3-3\rho^2+a^2(-3+4\rho^2))\arctan
\left(
\frac{\rho+\left(a-\sqrt{1-\rho^2}\right)\sqrt{\frac{1-x}{1+x}}}{\sqrt{a^2-1}}
\right)\Biggr.\\
&-&\sqrt{a^2-1}
\left(
\sqrt{1-\rho^2}+4 a x -8 a \rho^2 x- 2 \sqrt{1-\rho^2}x^2+4(1-4 a^2)\rho^3 \arccos{x}\right.\\
&+&2\rho\left(\left(-4a\sqrt{1-\rho^2}+x\right)\sqrt{1-x^2}+(-3+6a^2)\arccos{x}\right)\\
&+&\left.4\sqrt{1-\rho^2}(1-\rho^2+a^2(-1+4\rho^2))\log\left(
a+\sqrt{1-\rho^2} x +\rho\sqrt{1-x^2}\right)
\right)
\Biggl.\Biggr).
\end{eqnarray*}

We then have
$$\boxed{
P^{\G}\approx
\frac{1}{F^\beta\sigma i^{3/2}\sqrt{2\pi \tau}}
e^{-\frac{d(a,b_0)^2}{2s}+W0+WRevert}\left(1+\frac{4-5\rho^2}{24}\nu^2\tau\right)},$$
where the first-order term has been calculated thanks to the previous variations. The notations are given in the following.
\scriptsize
\begin{eqnarray*}
\zeta&=&\frac{\nu}{\sigma}\frac{f^{1-\beta}-F^{1-\beta}}{1-\beta}\\
i&=&\sqrt{\zeta^2-2\rho\zeta+1}\\
X_{0}&=&\frac{1}{\sqrt{1-\rho^2}}\left(\frac{f^{1-\beta}}{1-\beta}-\rho\frac{\sigma }{\nu}\right),\\
Y_{0}&=&\frac{\sigma}{\nu},\\
X_{end}&=&\frac{1}{\sqrt{1-\rho^2}}\left(\frac{F^{1-\beta}}{1-\beta}-\rho\frac{\sigma }{\nu}i\right),\\
Y_{end}&=&\frac{\sigma }{\nu}i,\\
l&=&\frac{(X_{end}^2+Y_{end}^2)-(X_0^2+Y_0^2)}{2(X_{end}-X_0)}\\
r&=&\sqrt{\left(\frac{X_{end}-X_{0}}{2}\right)^2+\left(\frac{Y_{end}-Y_{0}}{2(X_{end}-X_{0})}\right)^2+\frac{Y_{end}^2+Y_0^2}{2}}\\
a&=&\sqrt{1-\rho^2}\frac{l}{r}\\
f(q)&=&\frac{1}{2}\left(q\sqrt{1-q^2}+\arcsin{q}\right)\\
W0&=&\frac{-\beta}{2(1-\beta)\sqrt{1-\rho^2}} r^2\left(G\left(a,\rho,\frac{X_{end}-l}{r}\right)-G\left(a,\rho,\frac{X_0-l}{r} \right)\right)\\
WRevert&=&
-\rho \frac{\lambda}{\sqrt{1-\rho^2}} r^2\left(f\left(\frac{X_{end}-l}{r}\right)-f\left(\frac{X_0-l}{r} \right)\right)\\
&+&\frac{\rho \lambda\mu}{\sqrt{1-\rho^2}}(X_{end}-X_0)+
\frac{\lambda}{2}(Y_{end}^2-Y_0^2)-\lambda\mu(Y_{end}-Y_0)\\
d(a,b)&=&\log\left(\frac{i+\zeta-\rho}{1-\rho}\right)
\end{eqnarray*}
\normalsize

\boldmath
\subsection{Graphics}
\unboldmath

In the SABR case, here are the evolution of the transition probability and an example of a digital option that we got.

\begin{figure}[h]
   \begin{minipage}[c]{.46\linewidth}
      \includegraphics[width=7cm]{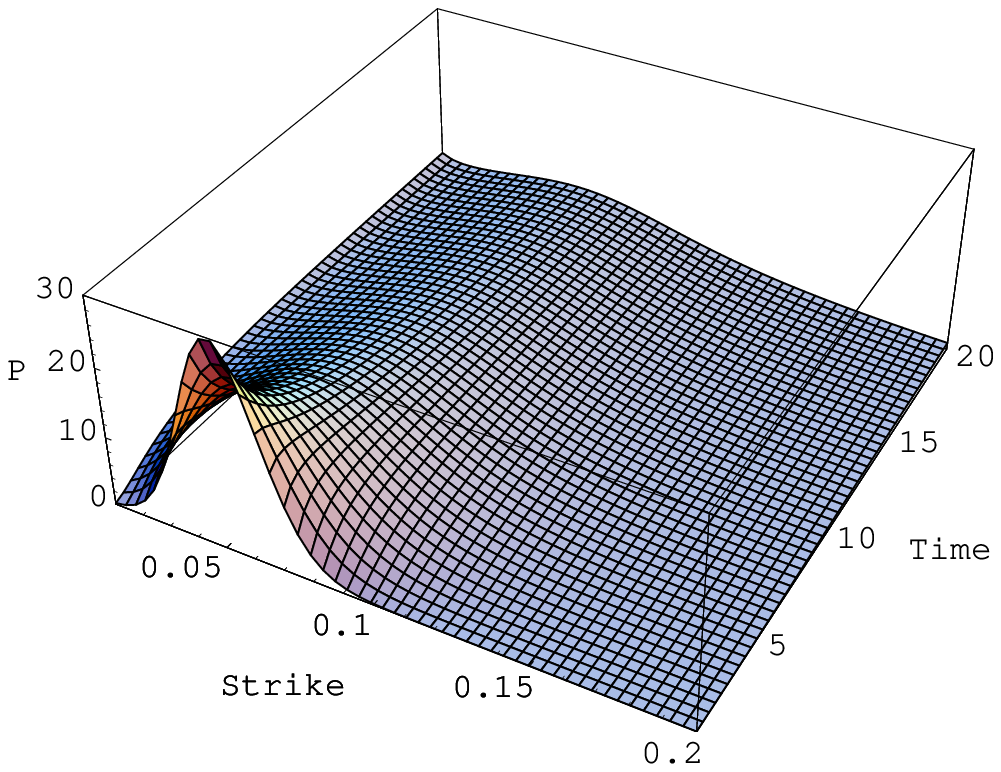}
   \end{minipage} \hfill
   \begin{minipage}[c]{.46\linewidth}
      \includegraphics[width=7cm]{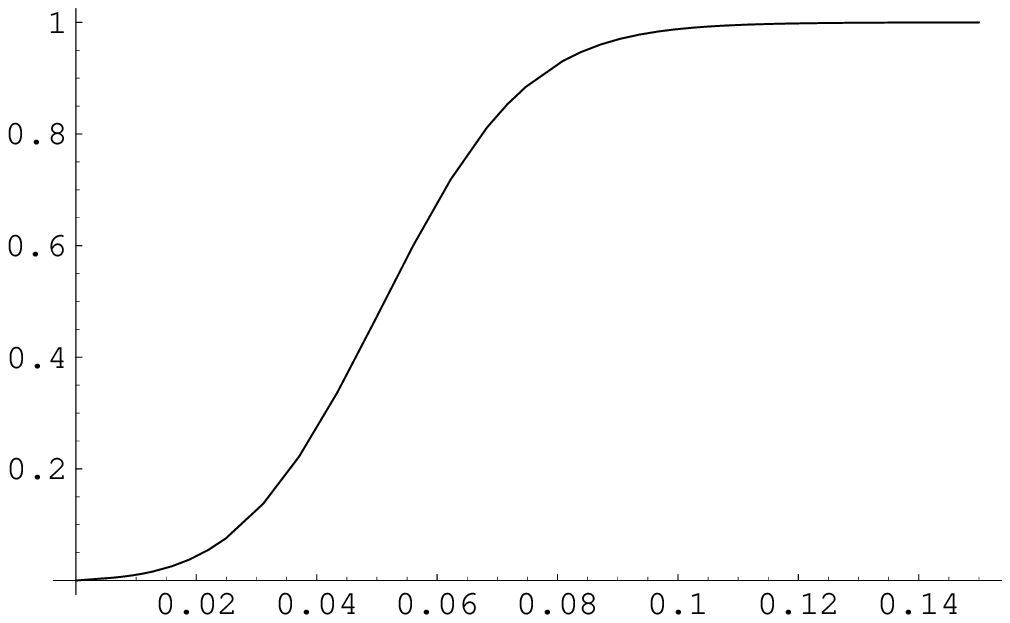}
   \end{minipage}
\end{figure}

The 1-order expression for $P$ seems to be relevant approximatively for $\tau\in[0,3]$, as shown in the following graphic : for $\tau>3$,
the 0-order expression is more stable : it remains aproximatively a distribution function, whereas the first-order does not.

\begin{figure}[h]
\begin{center}
      \includegraphics[width=9cm]{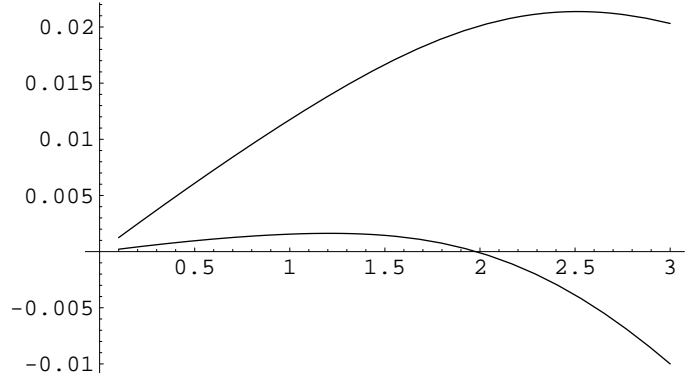}
\end{center}
\caption{$1-\int_\R P$ for the 0-order (up) and the 1-order (down) of $P$}
\end{figure}

\boldmath
\section{The implied volatility}
\unboldmath

\boldmath
\subsection{The local volatility : first order asymptotics}
\unboldmath

Local volatility can be defined as
$$\sigma_K(T,f,\sigma)^2\d T=\E{\left(\d F(T)\right)^2\mid F(T)=K}.$$
In the case of the $\delta$-model, (1) implies that
$$\sigma_K(T,f,\sigma)^2=C(K)^2\E{\left(\sigma(T)\right)^{2\delta}\mid F(T)=K}.$$

If we introduce the transition function
$G_{F,\Sigma}(\tau,f,\sigma)$, then a simple consideration such as
$P(A\mid B)=P(A\cap B)/P(B)$ (except that we reason
here with densities and zero-probability events) shows that
\begin{equation}
\sigma_K(\tau,f,\sigma)^2=\frac{C(K)^2\int_{0}^{\infty}\Sigma^{2\delta}
G_{K,\Sigma}(\tau,f,\sigma)\d \Sigma}{\int_{0}^{\infty}G_{K,\Sigma}(\tau,f,\sigma)\d \Sigma}=C(K)^2\frac{M_K(\tau,f,\sigma)}{P_K(\tau,f,\sigma)},
\end{equation}
where
\begin{itemize}
\item $P_K(\tau,f,\sigma)=\int_{0}^{\infty}G_{K,\Sigma}(\tau,f,\sigma)\d \Sigma$ is the marginal probability distribution;
\item $M_K(\tau,f,\sigma)=\int_{0}^{\infty}\Sigma^2
G_{K,\Sigma}(\tau,f,\sigma)\d \Sigma$ is the conditional moment of order $2\delta$.
\end{itemize}

At the leading order $P_K$ and $M_K$ have almost the same asymptotic expansion (as we can see with a saddle
point method) :
\begin{equation}
M_K^2(\tau,f,\sigma)\underset{\tau\to 0}{\sim}\Sigma_{min}^{2\delta}P_K(\tau,f,\sigma),
\end{equation}
with :

\begin{itemize}
\item $\Sigma_{min}=\nu Y^{\G}_{min}$;
\item $Y^{\G}_{min}=\nu^{\delta-1}Y^{\H}_{min}$;
\item $Y^{\H}_{min}$ the point of the straight line with equation
$$x=\frac{1}{\nu^{\delta-1}\sqrt{1-\rho^2}}\left(\int_{p}^K \frac{\d u}{c(u)}-\rho \nu^{\delta-1}y\right)$$
minimizing the distance to the point $\Phi(f,\sigma/\nu)$.
\end{itemize}

We can conclude with (1) and (2) that
$$\sigma_K(\tau,f,\sigma)=(\nu^\delta Y_{min}^{\H})^\delta c(K)+o(\tau).$$

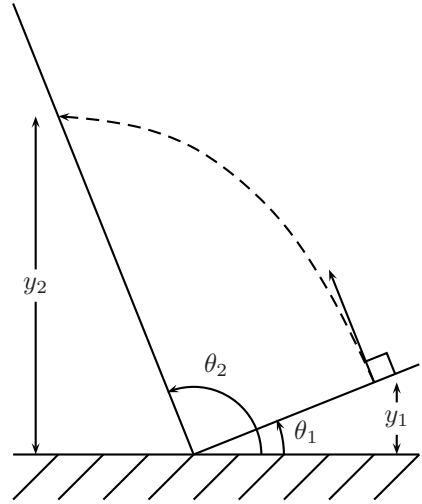
\begin{wrapfigure}[13]{r}{.5\linewidth}
\psset{unit=0.6cm}
\begin{center}
\begin{pspicture}(0,0)(9,9)
\psline(0,0)(9,0)
\psline(4,0)(0,10)
\psline(4,0)(9,2)
\psline(0,-1)(1,0)
\psline(1,-1)(2,0)
\psline(2,-1)(3,0)
\psline(3,-1)(4,0)
\psline(4,-1)(5,0)
\psline(5,-1)(6,0)
\psline(6,-1)(7,0)
\psline(7,-1)(8,0)
\psline(8,-1)(9,0)
\psline{->}(8,1.6)(7,4.1)
\angleplat{9}{2}{8}{1.6}{7}{4.1}{0.5}
\pscurve[linestyle=dashed]{->}(8,1.6)(7,3.6)(6,5.1)(5,6.1)(4,6.8)(3,7.2)(2,7.4)(1,7.5)
\psarc{->}(4,0){2}{0}{23}
\psarc{->}(4,0){1.5}{0}{113}
\psline{<->}(8.5,1.6)(8.5,0)
\psline{<->}(0.5,7.5)(0.5,0)
\rput(8.5,0.8){\psframebox*{$y_1$}}
\rput(0.5,3.75){\psframebox*{$y_2$}}
\rput(6.5,0.5){$\theta_1$}
\rput(4.5,2){$\theta_2$}
\end{pspicture}
\end{center}
\caption{The function $i_\delta(\theta_1,\theta_2)=y_1/y_2$}
\end{wrapfigure}

Unfortunately, we don't have a closed form for the distance from a point to a straight line in any $\delta$-space.
However, we can express $Y^\H_{min}$ in terms of a new 2-arguments function.

For this, imagine the following problem. In a $\delta$-space, a rocket is thrown perpendicularly to a straight line and crashes on another one.
Then we define
$$i_\delta(\theta_1,\theta_2)=\frac{y_1}{y_2}.$$
The function $g$ is well-defined because if $\gamma$ is a geodesic, so is $\lambda \gamma$

If we take $\theta_1$ and $\theta_2$ such as
\begin{itemize}

\item $\tan{\theta_1}=-\sqrt{1-\rho^2}/\rho$,

\item $\tan{\theta_2}=\sqrt{1-\rho^2}/(\nu/\sigma\ \int_{K}^{f}\d u/c(u)-\rho)$,

\end{itemize}

then
$$\sigma_K(\tau,f,\sigma)=\left(\sigma i_\delta(\theta_1,\theta_2)\right)^{\delta}c(K)+O(\tau).$$

If $\nu=0$ (no stochastic volatility), one can check that we have the expected value
$$\sigma_K(\tau,f,\sigma)=\sigma^{\delta}c(K)+O(\tau).$$

\boldmath
\subsection{From local to implied volatility}
\unboldmath

The previous calculation made it possible to transform the original stochastic volatility problem in a local volatility problem.
This is the reason why we can use a formula established by P. Hagan\footnote{{\sc Patrick S. Hagan and Diana E. Woodward,} {\it Equivalent Black Volatilities},
Applied Mathematical Finance 6, 147-157 (1999).}.\\

\fbox{
\begin{minipage}{0.9\textwidth}
{\bf Hagan's formula.} \it For a process $\d F=\alpha(t)c(F)\d W$, the log-normal implied volatility, at leading order in $\tau$, is given by
$$\sigma_B=a\frac{c(f_{av})}{f_{av}}\left[1+\frac{1}{24}\left(\frac{c''}{c}-2\left(\frac{c'}{c}\right)^2+\frac{2}{f_{av}^2}\right)(F_0-K)^2\right]+O(\tau)$$
with usually $f_{av}=(F_0+K)/2$ and $a=(1/\tau\ \int_{0}^{\tau}\alpha^2(t)\d t)^{1/2}$.\\
\rm
\end{minipage}
}
\vspace{0.3cm}

We now consider that for us $a$ can be replaced by $\sigma_K/c(f_{av})$. Indeed, for small times
\begin{eqnarray*}
\frac{\sigma_K^2}{c(f_{av})^2}&=&\frac{1}{\tau c(f_{av})^2}\E{\d F_\tau^2\mid F_\tau=K}\\
&=&\frac{1}{\tau c(f_{av})^2}\int_{0}^\tau\E{(\d F_t)^2\mid F_\tau=K}\\
&=&\frac{1}{\tau c(f_{av})^2}\E{\int_{0}^\tau\alpha^2(t)c(F_t)^2\d t\mid F_\tau=K}\\
\frac{\sigma_K^2}{c(f_{av})^2}&=&a^2
\end{eqnarray*}
if $f_{av}$ is defined by
$$\E{\int_{0}^\tau\alpha^2(t)c(F_t)^2\d t\mid F_\tau=K}=c(f_{av})^2\int_{0}^\tau\alpha^2(t)\d t.$$

Rather than $(f_0+K)/2$,
one should take, for small times,
$$f_{av}=\sqrt{\frac{K^2-f_0^2}{\log\left(\frac{K}{f_0}\right)}}.$$
Indeed this is the value that we got for a log-normal model ($c(F)=F$) (see appendix 3).\\

{\bf Remark.}  Pierre Henry-Labordère gave a different formula for the implied volatility
smile\footnote{{\sc Pierre Henry-Labordère,} {\it A General Asymptotic Implied Volatility for Stochastic Volatility Models,}
preprint.}.

\boldmath
\subsection{Smiles}
\unboldmath

The formulae above give us the exact asymptotic value of the smiles for $\tau=0$ :
$$\boxed{\sigma_B^{\tau\to 0}=(\sigma i_\delta(\theta_1,\theta_2))^\delta\frac{K^\beta}{f_{av}}\left(1+\frac{(1-\beta)(2+\beta)}{24}\left(\frac{K-f}{f_{av}}\right)^2\right)}.$$
with :
\begin{itemize}
\item $\tan{\theta_1}=-\sqrt{1-\rho^2}/\rho$;

\item $\tan{\theta_2}=\sqrt{1-\rho^2}/(\nu/\sigma\ \int_{K}^{f}\d u/c(u)-\rho)=\sqrt{1-\rho^2}/(\nu/\sigma\ \frac{K^(1-\beta)-f^(1-\beta)}{1-\beta}-\rho)$;
\item $f_{av}=\sqrt{\frac{K^2-f_0^2}{\log\left(\frac{K}{f_0}\right)}}$.
\end{itemize}

For different values of $\delta$, we get for instance the following smiles (Fig 3.4).

\begin{figure}[h]
\begin{center}
      \includegraphics[width=9cm]{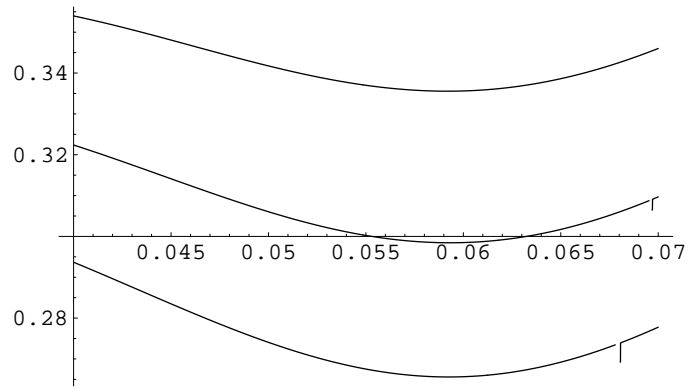}
\end{center}
\caption{Smiles from $\delta=0.5$ (down) to $\delta=0.7$ (up).}
\end{figure}

With all the previous results, one also can compute a first-order (in $\tau$) approximation of the smiles, which gives for example the following evolution (Fig 3.5).

\begin{figure}[h]
\begin{center}
      \includegraphics[width=9cm]{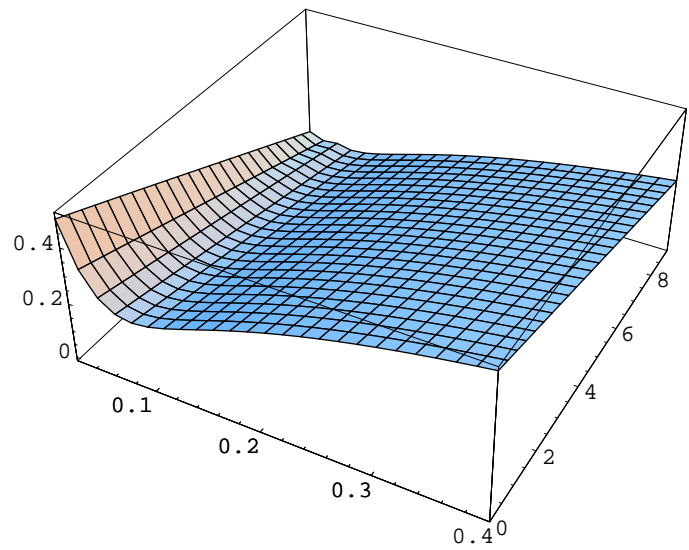}
\end{center}
\caption{First-order approximation for the evolution of a smile (SABR case).}
\end{figure}

\chapter{Appendix 1 : conditions for non-explosion of the volatility}{}{}

Let us first remind Feller's Test for Explosions.
We consider a process
$$\d X_t=a(X_t)\d t+b(X_t)\d W_t$$
on a segment $I=]l,r[$ of $\R$. We suppose that the functions $a$ and $b$ are Borel-measurable and follow the conditions of non-degeneracy
and local integrability :
\begin{itemize}

\item $(\forall x\in I)\  (b^2(x)>0)$;

\item $(\forall x\in I)\ (\exists \varepsilon>0)\ (\int_{x-\varepsilon}^{x+\varepsilon}\frac{1+|a(y)|}{b^2(y)}\d y<+\infty)$;

\end{itemize}

The scale function is defined, for a constant $c\in I$, as
$$p(x)=\int_c^x e^{-2\int_c^\xi\frac{a(\zeta) d \zeta}{b^2(\zeta)}}\d \xi.$$
One can then define the function
$$v(x)=\int_c^x p'(y)\int_c^y\frac{2\d z}{p'(z)b^2(z)}\d y.$$
The exit time from $I$ is
$$S=\inf\{t\geqslant 0\mid X_t\notin I\}.$$
We now can write the theorems we need\footnote{For a demonstration,
the reader can read {\sc Ioannis Karatzas, Steven E. Shreve,} {\it Brownian Motion and Stochastic Calculus,} Springer.}.\\

{\bf Proposition 1.} \it If a process $X_t$ follows the conditions above, with $X_0=x\in I$, then
\begin{itemize}
\item if $p(l^+)=-\infty$ and  $p(r^-)=+\infty$ then $P[S=\infty]=P[\sup_{0\leq t<+\infty}X_t=r]=P[\inf_{0\leq t<+\infty}X_t=l]=1$ :
the process is recurrent;
\item if $p(l^+)>-\infty$ and  $p(r^-)=+\infty$ then $P[\sup_{0\leq t<+\infty}X_t<r]=P[\lim_{t\to S}X_t=l]=1$;
\item if $p(l^+)=-\infty$ and  $p(r^-)<+\infty$ then $P[\sup_{0\leq t<+\infty}X_t>l]=P[\lim_{t\to S}X_t=r]=1$;
\item if $p(l^+)>-\infty$ and  $p(r^-)<+\infty$ then $P[\lim_{t\to S}X_t=l]=1-P[\lim_{t\to S}X_t=r]=\frac{p(r^-)-p(x)}{p(r^-)-p(l^+)}$;
\end{itemize}

\vspace{0.5cm}

{\bf Proposition 2 :  Feller's Test for Explosions.} \it If a process $X_t$ follows the conditions above, with $X_0\in I$, then
$$P[S=+\infty]=1\Leftrightarrow v(l^+)=v(r^-)=+\infty.$$

\rm

\vspace{0.5cm}

The case that interests us is
$$
\left\{
\begin{array}{ccc}
a(x)&=&\lambda(\mu-x)\\
b(x)&=&\nu\ x^\delta
\end{array}
\right.
$$
with $\nu>0$ and $\delta, \lambda,\mu\geqslant 0$. We will apply Feller's test successively for
$\delta\neq 1$ and $\delta\neq 1/2$, then $\delta=1$ and $\delta=1/2$.

\boldmath
\section{If $\delta\neq 1$ and $\delta\neq 1/2$}
\unboldmath

We have
\begin{eqnarray*}
p(x)&=&\int_c^x e^{-2\int_c^\xi\frac{\lambda(\mu-\zeta)d \zeta}{\nu^2\zeta^{2\delta}}}\d \xi,\\
p'(x)&=&e^{-2\int_c^x\frac{\lambda(\mu-\zeta)d \zeta}{\nu^2\zeta^{2\delta}}}\\
&=&e^{\frac{-2\lambda\mu}{\nu^2(1-2\delta)}(x^{1-2\delta}-c^{1-2\delta})+\frac{\lambda}{\nu^2(1-\delta)}(x^{2-2\delta}-c^{2-2\delta})}.
\end{eqnarray*}

We then have
$$\nu^{2\delta}v(x)=\int_{c}^x \underbrace{e^{\frac{-2\lambda\mu}{\nu^2(1-2\delta)}y^{1-2\delta}+\frac{\lambda}{\nu^2(1-\delta)}y^{2-2\delta}}
\int_c^y \frac{2\d z}{z^{2\delta}}e^{\frac{2\lambda\mu}{\nu^2(1-2\delta)}z^{1-2\delta}-\frac{\lambda}{\nu^2(1-\delta)}z^{2-2\delta}}}_{f(y)}\d y.$$

We will look for equivalents of $p$ or $f(y)$ in the following cases, to apply the proposition 1 or 2.

\subsubsection{If $\delta<1/2$}

In this case, one can write
$$
f(y)\underset{y\to 0}{\sim}
\int_c^y \frac{2\d z}{z^{2\delta}}e^{\frac{2\lambda\mu}{\nu^2(1-2\delta)}z^{1-2\delta}-\frac{\lambda}{\nu^2(1-\delta)}z^{2-2\delta}}
\underset{y\to 0}{\to}cst
$$
because $\frac{2}{z^{2\delta}}e^{\frac{2\lambda\mu}{\nu^2(1-2\delta)}z^{1-2\delta}-\frac{\lambda}{\nu^2(1-\delta)}z^{2-2\delta}}
\underset{z\to 0}{\sim}\frac{2}{z^{2\delta}}$, which is integrable in 0. So we have $v(0^+)<+\infty$ : by proposition 2 $P[S<+\infty]>0$.

\subsubsection{If $\delta>1$}

An analogous calculation as above gives $v(0^+)=v(+\infty)=+\infty$, so there is no explosion in a finite time.
However, if $\lambda=0$ (no mean reversion)
we have $p(+\infty)<+\infty$. If $\lambda\neq 0$, $p(0^+)=-\infty$ and $p(+\infty)=+\infty$, so by proposition 1 the process is recurrent.

\subsubsection{If $1/2<\delta<1$}

We now have
$$
f(y)\underset{y\to +\infty}{\sim}
e^{\frac{-2\lambda\mu}{\nu^2(1-2\delta)}y^{1-2\delta}+\frac{\lambda}{\nu^2(1-\delta)}y^{2-2\delta}}\times cst\geqslant cst
$$
for $y$ large enough. So $v(+\infty)=+\infty$. In $0^+$, we have, if $\lambda>0$,
$$
f(y)\underset{y\to 0}{\sim}e^{\frac{-2\lambda\mu}{\nu^2(1-2\delta)}y^{1-2\delta}}\times cst
$$
So, in this case $v(0^+)=+\infty$ : $P[S<+\infty]=0$. If $\lambda=0$, then
$$f(y)\underset{y\to 0}{\sim}\frac{1}{2\delta-1}\frac{1}{y^{2\delta-1}},$$
which is integrable in $0^+$. So, in this case, $P[S<+\infty]>0$.

Moreover, one can check that for $\lambda>0$, we
are in the first case of proposition 1, so the process is recurrent.

\section{If $\delta=1$ or $\delta=1/2$}

Simple calculations (where logarithms appear) show that, for these limit cases:
\begin{itemize}
\item
for $\delta=1$ there is no explosion and the process is recurrent;
\item
for $\delta=1/2$ there is no explosion if and only if $2\lambda\mu/\nu^2>1$; in such a case the process is recurrent.

\end{itemize}

\section{Summary}

The cases for which there is no explosion and\slash or limit for the process are :
\begin{itemize}

\item $\delta>1/2$ with mean reversion;

\item $\delta=1/2$ with $2\lambda\mu/\nu^2>1$.

\end{itemize}

From a practical point of view, we consider in our note the case $\delta\in[1/2,1]$ with no condition on $\lambda$, $\mu$ and $\nu$,
because our asymptotics are given for short times : the cases of explosions in the interval $[1/2,1]$ would be very pathological.

\chapter{Appendix 2 : $\delta$-geometry}{}{}

\it

In this part, we recall some useful results of
Riemannian geometry, and we apply it to the special case or the upper half plane ($y>0$) with the metric
\rm
$$\d s ^2=\frac{\d x^2+\d y^2}{y^{2\delta}},$$
\it
with $\delta\in\R ^+$. We will be particularly interested in the cases $\delta=1$ and $\delta=1/2$.
\rm

\boldmath
\section{Generalities about Riemannian spaces}
\unboldmath

In this section, we note $(g_{ij})_{1\leq i,j\leq n}$ the coefficients of a metric $G$, and we will use Einstein's summation conventions.

\boldmath
\subsection{Christoffel symbols}
\unboldmath

The Christoffel symbols are defined as
$$\Gamma_{ij}^k=\frac{1}{2}g^{kn}\left(\partial_i g_{nj}+\partial_j g_{ni}-\partial_n g_{ij}\right).$$

\boldmath
\subsection{Curvatures}
\unboldmath

The {\it Riemann tensor}, the {\it Ricci tensor} and the {\it scalar curvature} are successively defined by
$$
\left\{
\begin{array}{ccc}
{R^\alpha}_{\beta\gamma\delta}&=&\Gamma_{\beta\delta,\gamma}^\alpha-\Gamma_{\beta\gamma,\delta}^\alpha
+\Gamma_{\beta\delta}^\mu\Gamma_{\mu\gamma}^\alpha-\Gamma_{\beta\gamma}^\mu\Gamma_{\mu\delta}^\alpha\\
r_{\mu,\kappa}&=&{R^{\lambda}}_{\mu\lambda\kappa}\\
R&=&\frac{1}{2}g^{\mu\kappa}R_{\mu,\kappa}
\end{array}
\right.,
$$
where \og ,$\gamma$\fg denotes the partial derivative face to $\gamma$.
The {\it Gauss curvature} is defined in dimension 2 by
$$(R_{s,t}u)\cdot v=-R\ \mbox{vol}(s,t)\mbox{vol}(u,v),$$
that is to say, for $e_1$ and $e_2$ orthogonal vectors with norm 1,
$$R=\left(R_{e_1,e_2}e_2\right)\cdot e_1={R_{12}}^{21}.$$
It coincides with the scalar curvature, in dimension 2.

In dimension 2, one can show that
$$R=\frac{1}{\sqrt{g}}\left[\frac{\partial}{\partial_y}\left(\frac{\sqrt{g}}{g_{xx}}\Gamma_{xx}^y\right)
-\frac{\partial}{\partial_x}\left(\frac{\sqrt{g}}{g_{xx}}\Gamma_{xy}^y\right)\right],$$
with $g$ the determinant of $g$.

\boldmath
\subsection{Isometries}
\unboldmath

Let $\Phi$ be a function from $\G$ to $\H$, with respective metrics $G$ and $H$. Then $\Phi$ is said to be an isometry iff for every $a$ and $b$ in $\G$
$$d(a,b)_\G=d(\Phi(a),\Phi(b))_\H.$$

This global definition is equivalent to the following local condition : for every point $Z\in\G$,
$$G_Z=(\nabla \Phi)_Z^\top H_{\Phi(Z)}\nabla\Phi_Z.$$

\boldmath
\subsection{Geodesics}
\unboldmath

\boldmath
\subsubsection{Minimization of length}
\unboldmath

The length of a curve $c(t)$ in a Riemannian space $\G$, between $a$ and $b$, is defined by the integral
$$d(a,b)=\int \sqrt{<\p{c(t)},\p{c(t)}>_\G}\d t,$$
which is clearly not dependant of the parametrization.

For sufficiently regular spaces (for instance spaces with scalar curvature always negative, as in $\delta$-geometry), there exists an unique curve
$\gamma(t)$ minimizing the length between $a$ and $b$ along $\gamma$. This length, $l(a,b)$, is the {\it geodesic distance} between $a$ and $b$.
A non-trivial property is that this length coincides with the minimal energy between $a$ and $b$, that is to say the minimum along all ways of
$$\int <\p{c(t)},\p{c(t)}>_\G\d t,$$
under the normalization condition $|\p{c(t)}|_\G=1$.

\boldmath
\subsubsection{The parametric equation}
\unboldmath

Let $c(\tau)=[x(\tau),y(\tau)]$ be the position of a point at time $\tau$. Then an application of the Euler's perturbation formula gives
$$\nabla_{\p{c}}\p{c}=0.$$

In a coordinates system $(x_i)_{1\leq i\leq n}$, this can be written, for all $k\in\llbracket 1, n\rrbracket$
$$\pp{x}^k+\Gamma_{jl}^{k}\p{x}^j\p{x}^l=0.$$

\boldmath
\subsection{Killing fields}
\unboldmath

A field that preserves the metric is said to be a {\it Killing field}. In terms of the Lie derivative, this means that
$$\mathcal{L}_X g=0,$$
so for all fields $Y$ and $Z$
$$g(\nabla_YX,Z)+g(Y,\nabla_ZX)=0.$$

In particular,
if the metric is invariant along the direction $x_i$, then $\partial_{x_i}$ is a killing field. The previous equation then implies,
for a way $\gamma(t)$, $g(\nabla_{\p{\gamma(t)}}\partial_{x_i},\p{\gamma(t)})=0$. If $\gamma$ is a geodesic, we then have
\begin{eqnarray*}
\partial_t g(\partial_{x_i},\p{\gamma(t)})&=&g(\nabla_{\p{\gamma(t)}}\partial_{x_i},\p{\gamma(t)})+g(\partial_{x_i},\nabla_{\p{\gamma(t)}}\p{\gamma(t)})\\
&=&0+0\\
\partial_t g(\partial_{x_i},\p{\gamma(t)})&=&0.
\end{eqnarray*}

So, for a given geodesic,
one can associate to any Killing field an
invariant quantity along the geodesic :
$$g(\partial_{x_i},\p{\gamma(t)})=\mbox{cst}.$$

\boldmath
\subsection{Jacobi Fields}
\unboldmath

A field $\mathscr{J}(t)$ along a geodesic $\gamma(t)$ is said to be a {\it Jacobi Field} if and only if
$$\nabla^2_{\p{\gamma}}\mathscr{J}(t)+R(\mathscr{J}(t),\p{\gamma})\p{\gamma}=0,$$
where $R$ is the Riemann tensor.
This abstract definition becomes more intuitive thanks to the following proposition (see
Lang\footnote{{\sc Serge Lang,} {\it Fundamentals of Differential Geometry,} Springer.} for a demonstration).\\

\begin{wrapfigure}[8]{r}{.5\linewidth}
\psset{unit=0.8cm}
\begin{center}
\begin{pspicture}(2,0)(6,4)
\pscurve{->}(0,2)(1,2.5)(2.5,3)(4.5,3.5)(5.75,3.75)(7,4)
\pscurve{->}(0,2)(1,2.25)(2.5,2.5)(4.5,2.5)(5.75,2.25)(7,2)
\pscurve{->}(0,2)(1,2.75)(2.5,3.5)(4.5,4.25)(5.75,4.7)(7,5)
\psline[linewidth=0.5pt]{->}(1,2.5)(1,2.75)
\psline[linewidth=0.5pt]{->}(2.5,3)(2.5,3.5)
\psline[linewidth=0.5pt]{->}(4.5,3.5)(4.5,4.25)
\psline[linewidth=0.5pt]{->}(5.75,3.75)(5.75,4.7)
\rput(0,1.5){$t_0$}
\rput(7,3.5){$t_1$}
\rput(8,4){$\gamma(0,t)$}
\rput(8,5){$\gamma(s,t)$}
\end{pspicture}
\end{center}
\end{wrapfigure}

~~
{\bf Proposition.} \it
Let $\gamma\ :\ [a,b]\times,[t_0,t_1]\to \G$ be a variation of a geodesic $\gamma(t)$ such as for every $s\in[a,b]$ $\gamma(s,.)$ is a geodesic. Then
$$\mathscr{J}(t)=\partial_s\gamma(s,t)|^{s=0}$$
is a Jacobi Field. Furthermore, to every Jacobi Field along $\gamma$ one can associate such a variation.

\rm

\boldmath
\subsection{The Laplace Beltrami operator}
\unboldmath

The traditional Laplacian is defined as $\Delta(f)=\sum_{\mu}\partial_{\mu\mu}f$. In the euclidean case, this allows to write the relation
$\int \phi\Delta f\d v_{eucl}=-\int \nabla \varphi\nabla f\d v_{eucl}$ for any sufficiently regular function $\phi$ with compact support.

For any Riemannian space $\mathbb{M}$, with dimension $n$ and metric $M$, we want to find an expression of the Laplacian that verifies Stoke's theorem, that is to say
$$\int \varphi \Delta^\M f\d v^{\M}=-\int <\nabla \varphi,\nabla f>_\M \d v^{\M},$$
where $\d v^\M$ is the volume element of the Riemannian space (we know\footnote{{\sc Jûrgen Jost,} {\it Riemannian
Geometry and Geometric Analysis,} Springer Universitext, 2002.} that it is given by $\d v^\M=\sqrt{\det M}\ \d x^1\dots\d x^n$.)
Let $m$ be the determinant of the metric $\M$. As we have (we use Einstein's summation convention)
\begin{eqnarray*}
\int <\nabla \varphi,\nabla f>_\M \d v_{\M}&=&\int <\nabla \varphi,\nabla f>_\M \sqrt{m}\ \d x^1\dots \d x^n\\
&=&\int g^{ij}\frac{\partial \varphi}{\partial x^j}\frac{\partial f}{\partial x^i}\sqrt{m}\ \d x^1\dots\d x^n\\
\int <\nabla \varphi,\nabla f>_\M \sqrt{m}\ \d x^1\dots \d x^n&=&-\int\varphi \frac{1}{\sqrt{m}}\frac{\partial}{\partial x^j}\left(\sqrt{m}\ m^{ij}\frac{\partial f}{\partial x^i}\right)\sqrt{m}\ \d x^1\dots\d x^n,
\end{eqnarray*}
it is necessary and sufficient to define the Laplace Beltrami operator as
$$\Delta^\M=m^{-1/2}\partial_\mu\left(m^{1/2}m^{\mu \nu}\partial_\nu\right).$$

\begin{wrapfigure}[8]{r}{.4\linewidth}
\psset{unit=0.6cm}
\begin{center}
\begin{pspicture}(0,0)(6,3.2)
\psline{->}(0.5,4)(2.5,1)
\psline{->}(5.5,4)(3.5,1)
\psline{->}(1,4.5)(5,4.5)
\rput(3,4.5){\psframebox*[framearc=.3]{$\Phi$}}
\rput(1,2.5){\psframebox*[framearc=.3]{$\Delta^\G(f\circ\Phi)$}}
\rput(5,2.5){\psframebox*[framearc=.3]{$\Delta^\H(f)$}}
\rput(0.5,4.5){$\G$}
\rput(5.5,4.5){$\H$}
\rput(3,0.5){$\R $}
\pscurve[linestyle=dashed]{->}(5,4)(4,3.5)(3.2,2.3)(3,1)
\rput(4,3.5){\psframebox*[framearc=.3]{$f$}}
\end{pspicture}
\end{center}
\end{wrapfigure}

One of the advantages of this intrinsic formulation of the Laplacian is that it is invariant by isometry. More precisely,
if $\Phi$ is an isometry between two spaces $\G$ and $\H$, then for every function $f$ from $\H$ to $\R $ this diagram is commutative :
$$\Delta^\G(f\circ \Phi)=\Delta^\H (f)\circ\Phi.$$
Indeed, as an isometry conserves the scalar product (so the volume element, too), we have for every sufficiently regular function
$\varphi$ from $\G$ to $\R $ with compact support,
\begin{eqnarray*}
\int\Delta^\G(f\circ\Phi)\ \varphi\ \d v^\G&=&\int<\nabla(f\circ\Phi),\nabla \varphi>_\G\d v^\G\\
&=&\int<\nabla f,\nabla (\varphi\circ\Phi^{-1})>_\H\d v^\H\\
&=&\int\Delta^\H (f)\ \varphi\circ\Phi^{-1}\ \d v^\H\\
\int\Delta^\G(f\circ\Phi)\ \varphi\ \d v^\G&=&\int\Delta^\H (f)\circ\Phi\ \varphi\ \d v^\G
\end{eqnarray*}
As $\varphi$ is any sufficiently regular function with compact support, this gives the commutation relationship.

\boldmath
\section{A special case : the $\delta$-geometry}
\unboldmath

The  $\delta$-geometry is defined on the upper half plane ($y>0$) with the metric
\rm
$$\d s ^2=\frac{\d x^2+\d y^2}{y^{2\delta}},$$
with $\delta\in\R ^+$. We will represent it with the matrices
$$
G=
\left[
\begin{array}{cc}
\frac{1}{y^{2\delta}}&0\\
0&\frac{1}{y^{2\delta}}
\end{array}
\right], \ \
G^{-1}=
\left[
\begin{array}{cc}
y^{2\delta}&0\\
0&y^{2\delta}
\end{array}
\right].
$$

\rm

\boldmath
\subsection{The geodesics}
\unboldmath

After some calculation, the only Christoffel symbols that are not equal to 0 are :
\begin{itemize}
\item $\Gamma_{yx}^x=\Gamma_{xy}^x=1/2g^{xx}(\partial_y g_{xx})=-\delta/y$;
\item $\Gamma_{yy}^y=1/2g^{yy}(\partial_y g_{yy})=-\delta/y$;
\item $\Gamma_{xx}^y=1/2g^{yy}(-\partial_y g_{xx})=\delta/y$.
\end{itemize}

\boldmath
\subsubsection{Parametric equations}
\unboldmath

The parametric equation for geodesics is here
$$
\left\{
\begin{array}{ccc}
\pp{x}&=&\frac{2\delta}{y}\p{x}\p{y}\\
\pp{y}&=&\frac{\delta}{y}(\p{y}^2-\p{x}^2)
\end{array}
\right..
$$

Another (equivalent but more practical) differential system is the following,
$$
\left\{
\begin{array}{ccc}
\frac{\p{x}^2+\p{y}^2}{y^{2\delta}}&=&\mbox{cte1}\\
\frac{\p{x}}{y^{2\delta}}&=&\mbox{cte2}\\
\end{array}
\right..
$$

This system needs some explanation :
\begin{itemize}
\item the first equation expresses that $g(\p{\gamma},\p{\gamma})$ is constant;
\item the second one expresses that $\partial/\partial x$ is a killing field : the metric coefficients do not depend on $x$.
\end{itemize}
\vspace{0.3cm}

With these equations we can get the graph of these geodesics, leaving from $[0,1]$ with a speed $[1,0]$ and for $\tau \in[0,100]$ (Fig 5.1 and 5.2).

\begin{figure}[h]
\begin{center}
\includegraphics[width=5cm]{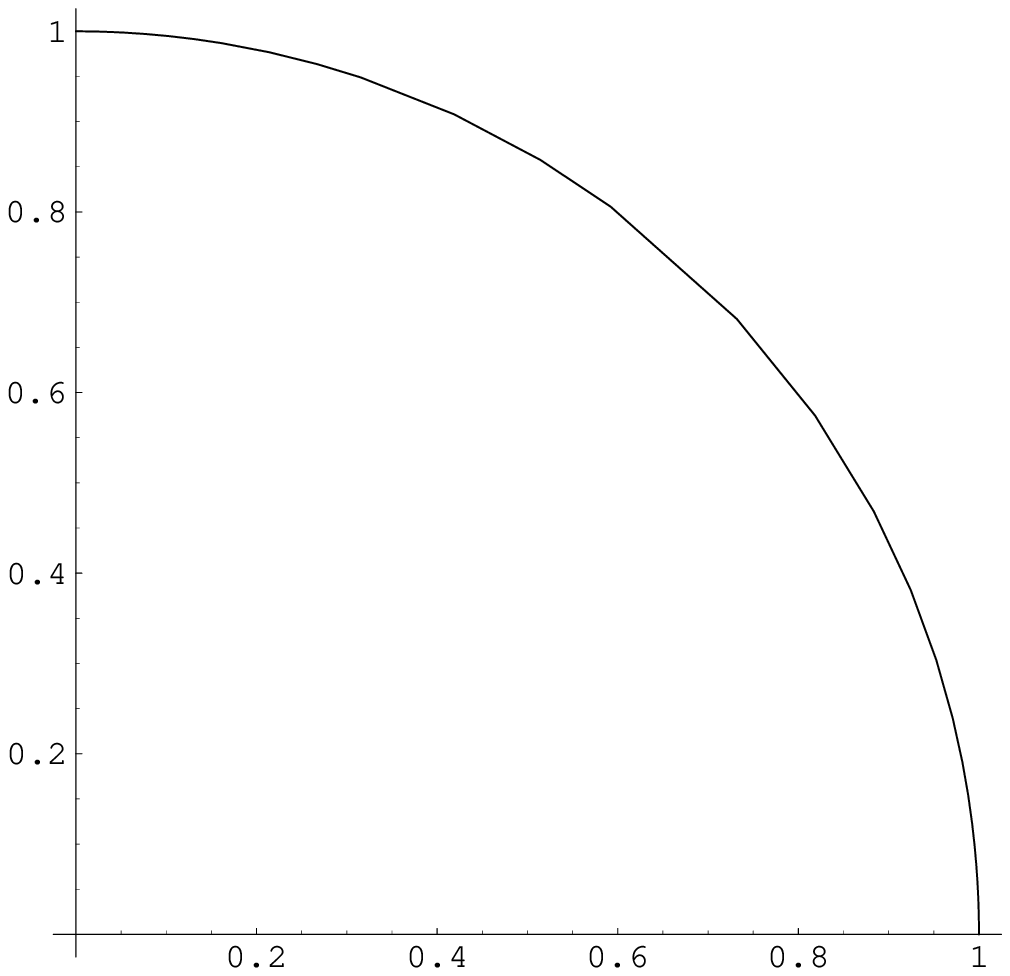}
\caption{$\delta=1$}
\end{center}
\end{figure}

\begin{figure}[h]
\begin{center}
\includegraphics[width=16cm]{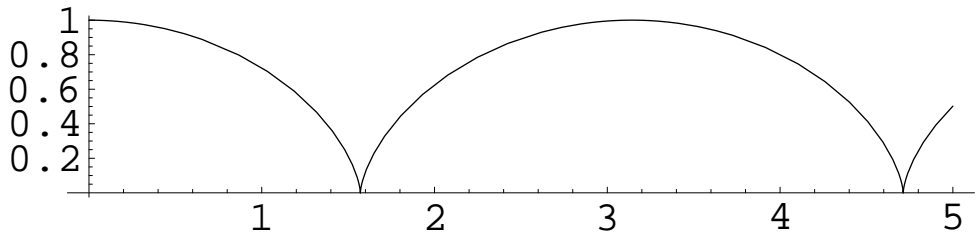}
\caption{$\delta=1/2$}
\end{center}
\end{figure}

\boldmath
\subsubsection{Explicit equation}
\unboldmath

Eliminating $\d t$ in the second system of the previous paragraph, we get (imposing a maximal possible $y$ equal to 1)
$$
\d x=\frac{y^\delta\d y}{\sqrt{1-y^{2\delta}}}
$$

An integration gives the following expression for $x(y)$, imposing $x(0)=0$ :

$$
x(y)=\frac{y^{\delta+1}}{\delta+1}\ \verb?Hypergeometric2F?\left(\frac{\delta+1}{2\delta},\frac{1}{2},\frac{3}{2}+\frac{1}{2d},y^{2\delta}\right).
$$
We get the following graphic for different values of $\delta$ (Fig 5.3).

\begin{figure}[h]
\begin{center}
\includegraphics[width=6cm]{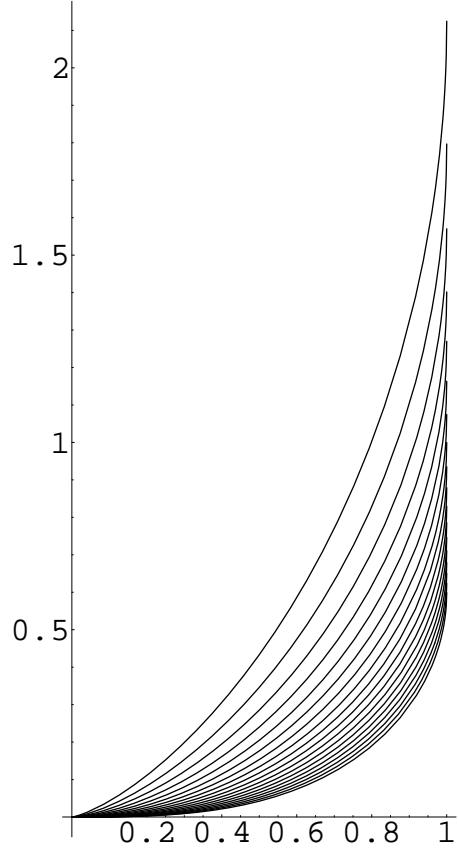}
\caption{$\delta\in[0.3,2]$}
\end{center}
\end{figure}

An interesting point is that for $\delta=1/n$ ($n\in\mathbb{N}$), Mathematica always finds an explicit formula.

\begin{center}
\begin{tabular}{cc}
\hline
\multicolumn{2}{c}{\bfseries Explicit formulae for $x(y)$}\\
\hline
$\delta$ & $x(y)$ \\
\hline
1   &  $1-\sqrt{1-y^2}$\\
1/2 &  $\arcsin{\sqrt{y}}-\sqrt{y-y^2}$\\
1/3 &  $2-2\sqrt{1-y^{3/2}}-\sqrt{y^{4/3}-y^2}$\\
1/4 &  $\frac{1}{2}\left(3\arcsin{y^{1/4}}-\sqrt{1-\sqrt{y}}(3+2\sqrt{y})y^{1/4}\right)$\\
\hline
\end{tabular}
\end{center}

\boldmath
\subsection{Curvatures}
\unboldmath

For our metric, the only non-zero curvatures are
$$
\left\{
\begin{array}{ccc}
{R^x}_{yxy}&=&-\frac{\delta}{y^2}\\
{R^x}_{yyx}&=&\frac{\delta}{y^2}\\
{R^y}_{xxy}&=&\frac{\delta}{y^2}\\
{R^y}_{xyx}&=&-\frac{\delta}{y^2}\\
\end{array}
\right.,
$$

$$
\left\{
\begin{array}{ccc}
r_{xx}&=&-\frac{\delta}{y^2}\\
r_{yy}&=&-\frac{\delta}{y^2}
\end{array}
\right.,
$$

$$R=y^{2\delta}\frac{\partial}{\partial y}\frac{\delta}{y}=-\delta y^{2\delta-2}.$$

\boldmath
\subsection{Homogeneity}
\unboldmath

Notice first that if ${\gamma(t)}_{t_0\leq t\leq t_1}$ is a line of the upper half plane (not necessarily a geodesic), then one can easily
know the length of $\lambda\gamma$ ($\lambda\in\R ^+_*$):
\begin{eqnarray*}
l(\lambda\gamma)&=&\int_{t_0}^{t_1}\sqrt{g(\p{\lambda\gamma},\p{\lambda\gamma})}\d t\\
&=&\int_{t_0}^{t_1}\sqrt{\frac{\p{\lambda x}^2+\p{\lambda y}^2}{(\lambda y)^{2\delta}}}\d t\\
&=&\lambda^{1-\delta}\int_{t_0}^{t_1}\sqrt{\frac{\p{x}^2+\p{y}^2}{(y)^{2\delta}}}\d t\\
l(\lambda\gamma)&=&\lambda^{1-\delta}l(\gamma)
\end{eqnarray*}

Another important homogeneity property is that if $\gamma$ is a geodesic, so is $\lambda \gamma$:
replacing $x$ and $y$ in
the parametric equations
defining the geodesics by $\lambda x$ and $\lambda y$ keeps the equalities true.

\boldmath
\subsection{Distance from a point to a straight line}
\unboldmath

For our metric, one can easily check that vertical lines are geodesics, so the distance from a point to a vertical is
gotten when the geodesic is perpendicular to the vertical axis.

So the homogeneity properties make it possible to calculate this length.

\psset{unit=1cm}
\begin{center}
\begin{pspicture}(-5,-1)(5,5)
\psplot{-3}{3}{9 x 2 exp neg add sqrt 0.5 mul}
\psplot{-5}{5}{25 x 2 exp neg add sqrt 0.5 mul}
\psline{->}(-6,0)(6,0)
\psline{->}(0,0)(0,5)
\psline(0,1.8)(-0.3,1.8)(-0.3,1.5)
\psline(0,2.8)(-0.3,2.8)(-0.3,2.5)
\psline[linestyle=dashed](-4,1.5)(0,0)
\psplot[linewidth=2pt]{-2.4}{0}{9 x 2 exp neg add sqrt 0.5 mul}
\psplot[linewidth=2pt]{-4}{0}{25 x 2 exp neg add sqrt 0.5 mul}
\rput(-4,1.9){$M$}
\rput(-2.4,1.3){$m$}
\rput(-0.4,3.7){$\Delta$}
\end{pspicture}
\end{center}

We have a point $M=(X,Y)$.
If the smallest geodesic is a \og reference\fg\ geodesic then one
can calculate the abscisse $x$ of the intersection of on the schema. Then
$$l(M,\Delta)=\left(\frac{X}{x}\right)^{\delta-1}l(x,\Delta).$$

\boldmath
\section{The first-order expansion at the origin.}
\unboldmath

To give $K_1$ at the origin $(x_0,y_0)$, we need to evaluate the terms in (2.4). In this equation $g$ can be replaced by $e^W\sqrt{d/Z(d)}=e^W h$.
First notice that $(e^W)_{(x_0,y_0)}=1$. Moreover, from the differential equation in the corollary 1, we have
$Z(d)=d+\frac{-R(x_0,y_0)}{3!}d^3+O(d^4)$ so
\begin{equation}
h=\sqrt{\frac{d}{Z(d)}}=\sqrt{\frac{d}{d+\frac{-R(x_0,y_0)}{3!}d^3+O(d^4)}}=1+\frac{1}{2}\frac{R(x_0,y_0)}{3!}\frac{x^2+y^2}{y_0^{2\delta}}+O(d^3)
\end{equation}
for near points $(x_0,y_0)$ and $(x_0+x,y_0+y)$.
We also need an expansion of $W$ up to the third order. This is quite technical but necessary. We consider the points
$a=(x_0,y_0)$ and $b=(x_0+x,y_0+y)$, and we note $W(a,b)=W(x,y)$. Then a classical Taylor expansion gives
$$W(x,y)=W_0+\left(\partial_x W\right)_0 x+\left(\partial_y W\right)_0 y
+\frac{1}{2}\left(\partial_{xx} W\right)_0 x^2+
\frac{1}{2}\left(\partial_{yy} W\right)_0 y^2+
\left(\partial_{xy} W\right)_0 x y+O(d^3).
$$

The evaluation of the partial derivatives with respect to $y$ is straightforward, because we know that the vertical lines are geodesics for the
$\delta-$space (cf appendix 2). More precisely,
$$W(0,y)=y \int_{0}^1 \frac{f_y(0,ty)\d t}{(y_0+ty)^{2\delta}},$$
so
$$\left(\partial_y W\right)_0=\frac{f_y(0,0)}{y_0^{2\delta}}$$
and
$$\left(\partial_{yy} W\right)_0=\frac{1}{{y_0}^{2\delta}}\left((\partial_y f_y)_0-2\frac{(f_y)_0}{y_0}\right).$$

The evaluation of the partial derivatives with respect to $x$ is more complicated, because the geodesic between $(x_0,y_0)$ and
$(x_0+x,y_0)$ is not a straight line anymore. Anyway, we will show that the calculation is the same as if it was a straight line indeed; to do so,
we will use the symmetry with respect to the line of points with abscise $x_0+x/2$ (cf appendix 2).

A simple look at the geodesics in $\delta$-geometry shows that the following schema is true : if $x$ is the length of the straight line, then
the length of the geodesic is $O(x)$ and the height is $O(x^2)$.

\psset{unit=1cm}
\begin{center}
\begin{pspicture}(0,-1)(11,3.5)
\psline{<->}(0,3)(10,3)
\psline{<->}(11,0)(11,2.5)
\psline[linewidth=4pt]{->}(1,0)(2.75,0)
\psline[linewidth=4pt]{->}(1,1)(2.75,2)
\psline[linestyle=dashed](0,0)(10,0)
\psline[linestyle=dotted](1,0)(1,1)
\psline[linestyle=dotted](2.75,0)(2.75,2)
\psline[linestyle=dashed](1,1)(3.5,1)
\pscurve(0,0)(1,1)(2.75,2)(5,2.5)(7.25,2)(9,1)(10,0)
\psarc(1,1){1}{0}{30}
\rput(5,3){\psframebox*{$x$}}
\rput(11,1.25){\psframebox*{$O(x^2)$}}
\rput(0,-0.5){$(x_0,y_0)$}
\rput(10,-0.5){$(x_0+x,y_0)$}
\rput(0,0){\textbullet}
\rput(10,0){\textbullet}
\rput(2.4,1.3){$\theta$}
\rput(2,-0.5){$\overrightarrow{\d l_2}$}
\rput(1.75,2){$\overrightarrow{\d l_1}$}
\rput(7.25,1.15){$\overrightarrow{f_1}$}
\rput(7.75,-0.75){$\overrightarrow{f_2}$}
\psline{->}(7.25,2)(8,1)
\psline{->}(7.25,0)(8.5,-0.5)
\end{pspicture}
\end{center}

Let us $W_1$ (resp $W_2$) be the work of the field $f$ along the geodesic (resp the straight line). We now show that
$$W_1=W_2+O(x^3).$$

The key point is that the geodesic has a symmetry axis, which implies $\int_0^x \tan\theta\d l_2=0$. This allows us to write
\begin{eqnarray*}
W_1-W_2&=&\int(\underbrace{\overrightarrow{f_1}-\overrightarrow{f_2})}_{O(x^2)}\cdot\overrightarrow{\d l_1}+\int\overrightarrow{f_2}\cdot(\overrightarrow{\d l_1}
-\overrightarrow{\d l_2})\\
&=&O(x^3)+\int\left[(f_2^y)_0+(\partial_xf_2^y)_0\l_2 +O(l_2^2)\right]\tan{\theta}\d l_2\\
&=&O(x^3)+\underbrace{\int(\partial_x f_2^y)_0\underbrace{\l_1}_{O(x)}\underbrace{\tan{\theta}}_{O(x)}\d l_2}_{O(x^3)}\\
W_1-W_2&=&O(x^3)
\end{eqnarray*}
Thus we have
$$(\partial_{xx}W_1)_0=(\partial_{xx}W_2)_0,$$
which can be easily calculated, like for the partial derivatives with respect to $y$. We finally get
$$(\partial_{xx}W_1)_0=\frac{1}{y_0^{2\delta}}(\partial_xf_x)_0.$$

We now can evaluate (2.4).
We have $h(x_0,y_0)=1$, so the equation (2.4) at the point $(x_0,y_0)$ becomes
$$(K_1)_0=\left[\left(\frac{1}{2}\Delta^\H+f\right)g\right]_0,$$
that is to say, for $g=e^Wh$,
\begin{eqnarray*}
(K_1)_0&=&\frac{1}{2}y_0^{2\delta}\Delta(e^Wh)_0+f(e^Wh)_0\\
&=&\frac{1}{2}y_0^{2\delta}\left[e^W\Delta+2\nabla e^W\cdot\nabla h+h\Delta e^W h\right]_0+\left[e^Wf h+h f e^W\right]_0\\
(K_1)_0&=&\frac{1}{2}y_0^{2\delta}\left[\underbrace{(\Delta h)_0}_{(1)}+2(\nabla e^W)_0\cdot\underbrace{(\nabla h)_0}_{(2)}
+\underbrace{|\nabla W|_0^2}_{(3)}+\underbrace{(\Delta W)_0}_{(4)}\right]+\underbrace{(f h)_0}_{(5)}+\underbrace{(f W)_0}_{(6)}
\end{eqnarray*}

We now have to evaluate each term :
\begin{itemize}
\item from (2.5) one can easily conclude that $(1)=R(x_0,y_0)/(3y_0^{2\delta})$ and $(2)=(5)=0$;
\item the first-order approximation for $W$ gives
$(3)={|\overrightarrow{f_0}|}^2/{y_0}^{4\delta}$ and $(6)={|\overrightarrow{f_0}|}^2/{y_0}^{2\delta}$;
\item the second-order evaluation of $W$ gives
$$(4)=\frac{1}{{y_0 }^{2\delta}}\left((\operatorname{div} f)_0-2\frac{(f_y)_0}{y_0}\right).$$
\end{itemize}

We then finally have
$$(K_1)_0=\frac{-\delta y_0^{2\delta-2}}{6}+\frac{(\operatorname{div} f)_0}{2}-\frac{(f_y)_0}{y_0}+\frac{3}{2}\frac{|f_0|^2}{y_0^{2\delta}}.$$

\boldmath
\section{Two new functions}
\unboldmath

Thanks to the above homogeneity properties, we only need to calculate the distances and the Jacobi fields on the standard geodesic
(symmetric with respect to the ordinates axis and $y[0]=1$). Here is the natural way to do all the necessary calculations on the standard geodesic :
\begin{itemize}
\item the function $\mbox{Sind}$ gives the position that minimizes the distance from a point to a line;
\item the distance function is specially easy to compute on the standard geodesic : we have a formula with only the coordinates of the end points;
\item then, to compute the Jacobi fields,we use the fact that it is a 2-dimensional linear space.
\end{itemize}

\boldmath
\subsection{The function $\sind$}
\unboldmath

We write $x(y)$the function such as
$$
\left\{
\begin{array}{ccc}
x(1)&=&0\\
\frac{\d x}{\d y}&=&\frac{-y^{2\delta}}{\sqrt{1-y^{2\delta}}}
\end{array}
\right..
$$

\boldmath
\subsubsection{Geometric definition of function $\sind$}
\unboldmath

The intuitive definition of function $\sind$ is given by the following drawing.

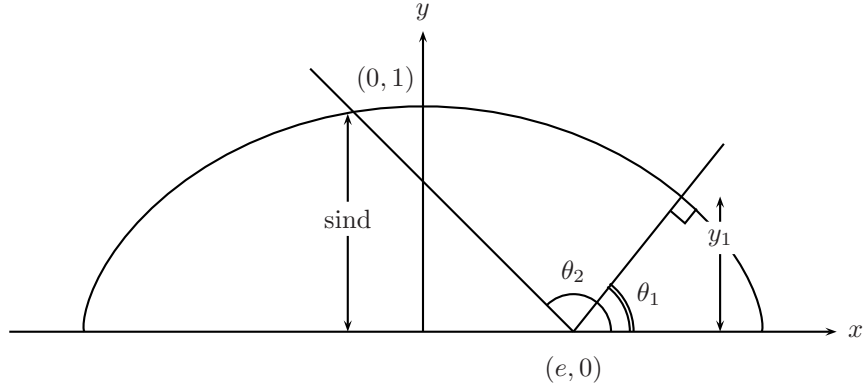
\begin{figure}[h]
\psset{unit=0.5cm}
\begin{center}
\begin{pspicture}(-5,0)(5,8.5)
\parametricplot{-180}{180}{t 60 div t sin add 2 div 6 mul 1 t cos add 2 div 6 mul}
\psline{->}(-11,0)(11,0)
\psline{->}(0,0)(0,8)
\psline(4,0)(8,5)
\psline(4,0)(-3,7)
\angleplat{4}{0}{6.9}{3.6}{8.3}{2.4}{0.5}
\psarc(4,0){1.5}{0}{53}
\psarc(4,0){1.6}{0}{53}
\psarc(4,0){1}{0}{135}
\psline{<->}(7.9,3.6)(7.9,0)
\psline{<->}(-2,5.8)(-2,0)
\rput(6,1){$\theta_1$}
\rput(4,1.6){$\theta_2$}
\rput(7.9,2.5){\psframebox*{$y_1$}}
\rput(-2,3){\psframebox*{\sind}}
\rput(-1,6.7){$(0,1)$}
\rput(4,-1){\psframebox*{$(e,0)$}}
\rput(11.5,0){$x$}
\rput(0,8.5){$y$}
\end{pspicture}
\end{center}
\caption{The geometric view of function $\sind$}
\end{figure}

\boldmath
\subsubsection{Analytical definition of function $\sind$}
\unboldmath

For $x>0$, we have
$$\frac{\d x}{\d y}=\frac{-y^{2\delta}}{\sqrt{1-y^{2\delta}}},$$
so
$$y_1=\left(\sin{\theta_1}\right)^{\frac{1}{\delta}}.$$

A simple calculation then gives
$$
e=
\left\{
\begin{array}{ccc}
\mbox{if $0< \theta_1\leq \frac{\pi}{2}$,}&& x(y1)-\frac{y1}{\tan{\theta_1}}\\
\mbox{if $\frac{\pi}{2}\leq \theta_1< \pi$,}&& -x(y1)-\frac{y1}{\tan{\theta_1}}
\end{array}
\right..
$$

Then the function $\sind(\delta,\theta_1,\theta_2)$ is the unique solution of the following equation :
\begin{equation}
\left\{
\begin{array}{ccc}
\mbox{if $e+\frac{1}{\tan{\theta_2}}\geq 0$,}&& e+\frac{\sind}{\tan{\theta_2}}=x(\sind)\\
\mbox{if $e+\frac{1}{\tan{\theta_2}}\leq 0$,}&& e+\frac{\sind}{\tan{\theta_2}}=-x(\sind)
\end{array}
\right..
\end{equation}

\boldmath
\subsubsection{Differential properties of function $\sind$}
\unboldmath

The equation (1) defines $\sind$ in an implicit way. One can now get the differentials of $\sind$, differentiating (1). The equations should be different
in 4 zones of $[0,\pi]\times[0,\pi]$, but we just have 2 such zones in reality.

$$
\left\{
\begin{array}{ccc}
\mbox{if $e+\frac{1}{\tan{\theta_2}}\geq 0$,}&& \left(\frac{\sind^\delta}{\sqrt{1-\sind^{2\delta}}}+\frac{1}{\tan{\theta_2}}\right)\partial_{\theta_1}\sind=
\left(\frac{1}{\delta}-1\right)\left(\sin{\theta_1}\right)^{\frac{1}{\delta}-2}\\
\mbox{if $e+\frac{1}{\tan{\theta_2}}\leq 0$,}&& \left(-\frac{\sind^\delta}{\sqrt{1-\sind^{2\delta}}}+\frac{1}{\tan{\theta_2}}\right)\partial_{\theta_1}\sind=
\left(\frac{1}{\delta}-1\right)\left(\sin{\theta_1}\right)^{\frac{1}{\delta}-2}
\end{array}
\right.
$$

$$
\left\{
\begin{array}{ccc}
\mbox{if $e+\frac{1}{\tan{\theta_2}}\geq 0$,}&& \left(\frac{\sind^\delta}{\sqrt{1-\sind^{2\delta}}}+\frac{1}{\tan{\theta_2}}\right)\partial_{\theta_2}\sind=
\frac{\sind}{\left(\sin{\theta_2}\right)^2}\\
\mbox{if $e+\frac{1}{\tan{\theta_2}}\leq 0$,}&& \left(-\frac{\sind^\delta}{\sqrt{1-\sind^{2\delta}}}+\frac{1}{\tan{\theta_2}}\right)\partial_{\theta_2}\sind=
\frac{\sind}{\left(\sin{\theta_2}\right)^2}
\end{array}
\right.
$$

\boldmath
\subsubsection{Examples of function $\sind$}
\unboldmath

One can check that for $\delta=1$ (geodesics are circles) the $\sind(\theta_1,\theta_2=\sin(\theta_1)/\sin(\theta_2))$.

\begin{figure}[h]
\begin{center}
\includegraphics[width=6cm]{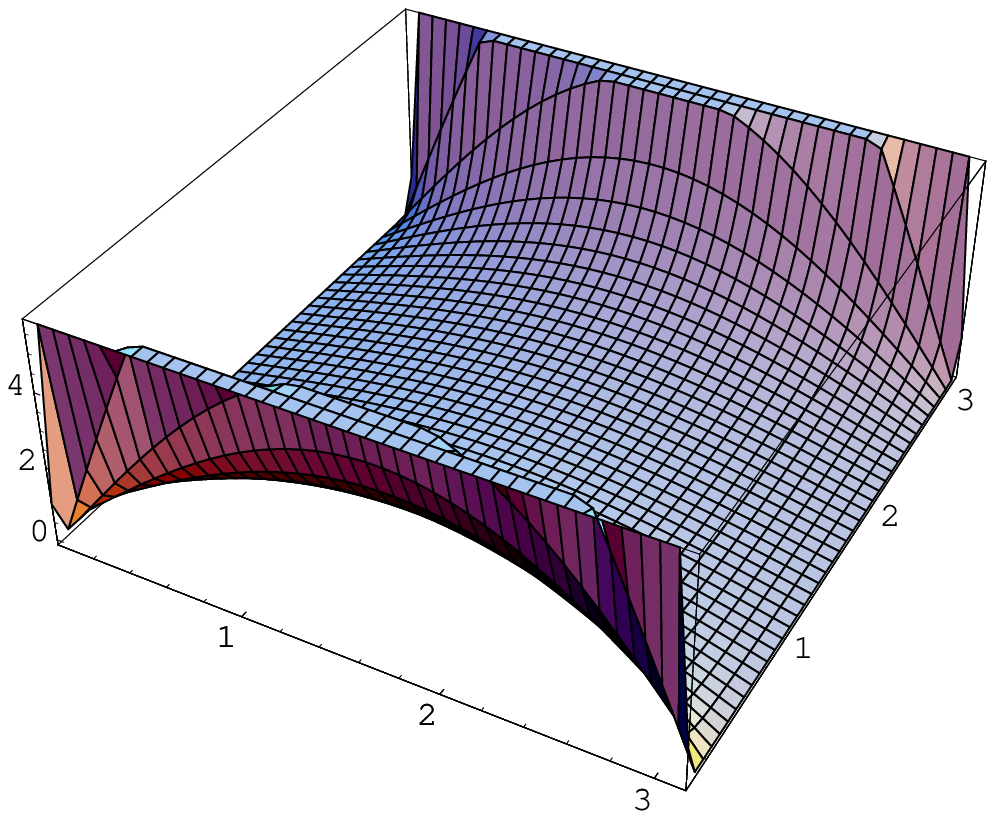}
\caption{$\delta=1$}
\end{center}
\end{figure}

\begin{figure}[h]
\begin{center}
\includegraphics[width=6cm]{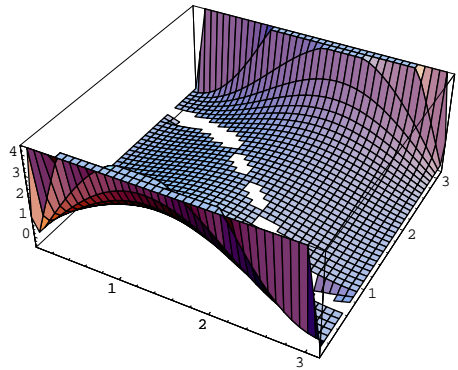}
\caption{$\delta=1/2$}
\end{center}
\end{figure}

The undefined values appearing on the frontier above is due to the difficulties Mathematica has to solve the equation (2.1) for
$$e+\frac{1}{\tan{\theta_2}}\approx 0.$$

\boldmath
\subsection{The distance}
\unboldmath

Thanks to the homogeneity properties, we only need to know the distance between the point (0,1) and any other point of the standard geodesic.
This distance is exactly
\begin{eqnarray*}
d((0,1)\to(x_1,y_1))&=&\int\sqrt{\frac{\d x^2+\d y^2}{y^{2\delta}}}\\
&=&\int_{y_1}^1\frac{\d y}{\sqrt{y^{2\delta}(1-y^{2\delta})}}\\
d((0,1)\to(x_1,y_1))&=&\frac{x_1-y_1^{1-\delta}\sqrt{1-y_1^{2\delta}}}{1-\delta}
\end{eqnarray*}
thanks to links between hypergeometric functions.

\boldmath
\subsection{The Jacobi field}
\unboldmath

The equation giving the Jacobi field on any geodesic is linear, so it is sufficient to evaluate it on the standard geodesic.
In order to evaluate the Jacobi field on the standard geodesic with any initial point and initial increase,
we just need to calculate two specific geodesics and then use linear combinations.
The most natural initial conditions are :
\begin{itemize}
\item $Z(0,1)=0$ and $\p{Z}(0,1)=1$;
\item $Z(0,1)=1$ and $\p{Z}(0,1)=0$;
\end{itemize}

The differential equation giving the evolution of a Jacobi field (we just need to know its length because it remains orthogonal to the geodesic) is
\begin{eqnarray*}
&&\pp{Z}+K\ Z=0\\
&\Leftrightarrow&\pp{Z}-\delta y^{2\delta-2} Z=0.
\end{eqnarray*}

For $\delta=1$, the equation above has the evident solution $Z(d)=A \cosh{d}+B\sinh{d}$.
For $\delta\neq 1$, we need to reformulate the equation above in terms of the coordinate $y$. We get
$$(1-y^{2\delta})\partial_{yy}Z+\delta\frac{1-2y^{2\delta}}{y}\partial_yZ-\frac{\delta}{y^2}Z=0.$$

The problem is that this equation is not lipschitzian for $y=1$. This is the reason why,
to get a great approximation of the solution, we use the initial conditions
$$
\left\{
\begin{array}{ccc}
Z_1[1-\varepsilon]&=&0\\
\p{Z_1}[1-\varepsilon]&=&1
\end{array}
\right.,\ \
\left\{
\begin{array}{ccc}
Z_2[1-\varepsilon]&=&1\\
\p{Z_2}[1-\varepsilon]&=&0
\end{array}
\right.
$$
with $\varepsilon$ small enough (typically $\varepsilon=1/100$).
If we take the $x$-coordinates, we then just need to take :
\begin{itemize}
\item if $x>0$, $Z_1[x]:=Z_1[y[x]]$, if $x<0$, $Z_1[x]:=-Z_1[y[x]]$;
\item if $x>0$, $Z_2[x]:=Z_2[y[x]]$, if $x<0$, $Z_2[x]:=Z_2[y[x]]$.
\end{itemize}

Then, linear combinations of $Z_1$ and $Z_2$ give any Jacobi Field.

\newpage
\boldmath
\section{Variation formulas for the $\delta$-model}
\unboldmath

The reader who does not know anything about the following variation calculations
may read the famous books
by Milnor\footnote{{\sc J. Milnor,} {\it Morse Theory}, Annals of mathematics studies, Princeton university press.}
or Lang\footnote{{\sc Serge Lang,} {\it Fundamentals of Differential Geometry,} Springer.}.

\boldmath
\subsection{The first variation formula : minimization implies orthogonality}
\unboldmath

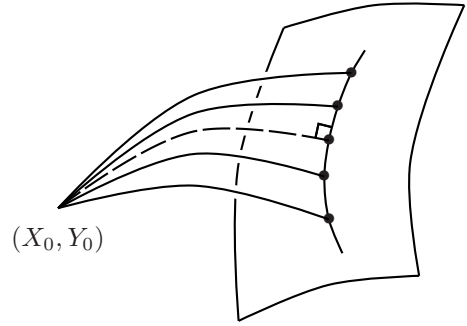
\begin{wrapfigure}[18]{r}{.5\linewidth}
\psset{unit=0.6cm}
\begin{center}
\begin{pspicture}(0,1)(9,7)

\pscurve(4,0.5)(6,1.3)(8,1.5)
\pscurve(8,1.5)(7.8,4)(9,7.5)
\pscurve(9,7.5)(7,7.5)(5,7)
\pscurve(5,7)(4,4)(4,0.5)

\pscurve[linecolor=white, linewidth=5pt](0,3)(3,3.5)(6,2.75)
\pscurve[linecolor=white, linewidth=5pt](0,3)(3,4.2)(5.9,3.7)
\pscurve[linecolor=white, linewidth=5pt](6,4.5)(3,4.7)(0,3)
\pscurve[linecolor=white, linewidth=5pt](0,3)(3,5.1)(6.2,5.25)
\pscurve[linecolor=white, linewidth=5pt](0,3)(3,5.5)(6.5,6)
 \rput(0,2.3){$(X_0,Y_0)$}

\psline(5.7,4.55)(5.77,4.87)(6.05,4.8)
\pscurve(6.3,2)(6,2.75)(5.9,3.7)(6,4.5)(6.2,5.25)(6.5,6)(6.8,6.5)
\rput(6,2.75){\textbullet}
\rput(5.9,3.7){\textbullet}
\rput(6,4.5){\textbullet}
\rput(6.2,5.25){\textbullet}
\rput(6.5,6){\textbullet}
\pscurve(0,3)(3,3.5)(6,2.75)
\pscurve(0,3)(3,4.2)(5.9,3.7)
\pscurve[linestyle=dashed,dash=10pt 2pt](6,4.5)(3,4.7)(0,3)
\pscurve(0,3)(3,5.1)(6.2,5.25)
\pscurve(0,3)(3,5.5)(6.5,6)
\psline(5.7,4.55)(5.77,4.87)(6.05,4.8)
\end{pspicture}
\end{center}
\caption{Geodesics between a point and a hypersurface in a curved space : orthogonality at the critical point.}
\end{wrapfigure}

We consider sufficient conditions of regularity (uniqueness of the geodesic between two points etc).
Then the distance from a point $Z_0$ to a line is minimal in a point $Z_{min}$
(with vector along the geodesic $\p{\gamma}(Z_{min})$ and along the line directed by $v(Z_{min})$) if and only if
$$<\p{\gamma}(Z_{min}),v(Z_{min})>_\H=0.$$
As $\H$ is locally euclidean, this is equivalent to the usual orthogonality. Let us justify this relationship.
For a variation $\gamma(s,t)$ (where $t$ represents the proper time, that is to say the length), if $\gamma(s,\cdot)$ is a geodesic for every
$s$, then the length and the energy are equal, that is to say
$$d(s)=\frac{1}{2}\int \sqrt{<\partial_t\gamma(s,t),\partial_t\gamma(s,t)>_\H}\d t=\frac{1}{2}\int <\partial_t\gamma(s,t),\partial_t\gamma(s,t)>_\H\d t.$$
This is a simple consequence of the definition of the proper time : $<\partial_t\gamma(s,t),\partial_t\gamma(s,t)>_\H=1$.
We then can write
\begin{eqnarray}
\partial_s d(s)&=&\frac{1}{2}\int \partial_s<\partial_t\gamma(s,t),\partial_t\gamma(s,t)>_\H\d t\nonumber\\
&=&\frac{1}{2}\int \left(<\nabla_{\partial_s}\partial_t\gamma(s,t),\partial_t\gamma(s,t)>_\H+<\partial_t\gamma(s,t),\nabla_{\partial_s}\partial_t\gamma(s,t)>_\H\right)\d t\\
&=&\frac{1}{2}\int \left(<\nabla_{\partial_t}\partial_s\gamma(s,t),\partial_t\gamma(s,t)>_\H+<\partial_t\gamma(s,t),\nabla_{\partial_t}\partial_s\gamma(s,t)>_\H\right)\d t\\
&=&\int <\nabla_{\partial_t}\partial_s\gamma(s,t),\partial_t\gamma(s,t)>_\H\d t\nonumber\\
&=&\int \left(\partial_t<\partial_s\gamma(s,t),\partial_t\gamma(s,t)>_\H-
<\partial_s\gamma(s,t),\nabla_{\partial_t}\partial_t\gamma(s,t)>_\H \right)\d t\\
&=&\int \partial_t<\partial_s\gamma(s,t),\partial_t\gamma(s,t)>_\H\d t\\
&=&<\partial_s\gamma(s,t),\partial_t\gamma(s,t)>_\H|_{t=0}^{t=d}\nonumber\\
\partial_s d(s)&=&<\partial_s\gamma(s,t),\partial_t\gamma(s,t)>_\H|^{t=d}.
\end{eqnarray}
The calculation above requires some explanation :
\begin{itemize}
\item (2.1) : we use the well-known relation $\partial_x<u,v>=<\nabla_{\partial_x}u,v>+<u,\nabla_{\partial_x}v>$;
\item (2.2) : the connection $\nabla$ is torsion-free;
\item (2.3) : we also use the relationship $\partial_x<u,v>=<\nabla_{\partial_x}u,v>+<u,\nabla_{\partial_x}v>$;
\item (2.4) : $\gamma$ is a geodesic for every $s$, so $\nabla_{\partial_t}\partial_t\gamma(s,t)=0$;
\item (2.5) : the initial point does not move with the variation, so $\partial_s\gamma(s,t)|_{t=0}=0$.
\end{itemize}

\boldmath
\subsection{Second variation for any strike}
\unboldmath

In the following, we will have to calculate partial derivatives with respect to $s$, so for simplicity we will write $\eta=\partial_s$.
We will then often use the following expression for the connection with direction $\eta$,
\begin{equation}
\nabla_\eta u=
\left[
\begin{array}{c}
\partial_s u_x\\
\partial_s u_y
\end{array}
\right]
+
\left[
\begin{array}{c}
\Gamma^x_{xy}\eta_x u_y+\Gamma^x_{yx}\eta_y u_x\\
\Gamma^y_{xx}\eta_x u_x+\Gamma^y_{yy}\eta_y u_y
\end{array}
\right]
=
\left[
\begin{array}{c}
\partial_s u_x\\
\partial_s u_y
\end{array}
\right]
+
\frac{\delta}{y}
\left[
\begin{array}{c}
-\cos{\theta_1}u_y -\sin{\theta_1}u_x\\
\cos{\theta_1}u_x- \sin{\theta_1}u_y
\end{array}
\right],
\end{equation}
where $\theta_1$ is the angle between the straight line and the abscises axis.

To calculate the second derivative, we derive it from the expression (2.5),
\begin{eqnarray}
\partial_{ss}d(s)&=&\partial_s<\partial_s\gamma(s,t),\partial_t\gamma(s,t)>_\H|^{t=d}\nonumber\\
&=&<\nabla_{\partial_s}\partial_s\gamma(s,t),\partial_t\gamma(s,t)>_\H|^{t=d}+
<\partial_s\gamma(s,t),\nabla_{\partial_s}\partial_t\gamma(s,t)>_\H|^{t=d}\nonumber\\
&=&<\nabla_{\partial_s}\partial_s\gamma(s,t),\partial_t\gamma(s,t)>_\H|^{t=d}+
<\partial_s\gamma(s,t),\nabla_{\partial_t}\partial_s\gamma(s,t)>_\H|^{t=d}\nonumber\\
\partial_{ss}d(s)&=&\underbrace{<\nabla_{\eta}\eta,\partial_t\gamma(s,t)>_\H|^{t=d}}_{(1)}+
\underbrace{\frac{1}{2}\partial_t<\eta,\eta>_\H|^{t=d}}_{(2)}
\end{eqnarray}
where we use that the connection is torsion-free, and the relationship of derivation of the scalar product.
Let us evaluate each of both terms.
\begin{itemize}

\item We first have to calculate the covariant derivative
$$\nabla_\eta\eta=
\frac{\delta}{y}
\left[
\begin{array}{c}
-2\cos{\theta_1}\sin{\theta_1}\\
\cos^2{\theta_1}-\sin^2{\theta_1}
\end{array}
\right]
$$
at the point $Z_{min}$
where $\theta_1$ is the angle between the abscises axis and $\mathbb{D}$. Moreover,
$$
\p{\gamma}=
\pm
y^\delta
\left[
\begin{array}{c}
\sin{\theta_1}\\
-\cos{\theta_1}
\end{array}
\right]
$$
where the sign + or - depends on which side the point $Z_0$ is with respect to $\mathbb{D}$. So we have
$$(1)=\pm \frac{\delta}{Y_{min}^{\delta+1}}\cos\theta_1.$$

\item As parallel transport keeps orthogonality, we have
$\eta|^{s=0}=\frac{Z(t)}{Z(d)} \overrightarrow{u}$, where $\overrightarrow{u}$ is the orthogonal of the speed vector
with norm 1 (no matter its direction).
So we have
$$(2)=\frac{\p{Z}(d)}{Z(d)Y_{min}^{2\delta}}.$$
\end{itemize}

To sum up, we have
$$\frac{\d^2}{\d y^2}d(Y_{min})=
\frac{1}{\sin^2\theta_1}\left(\frac{\p{Z}(d)}{Z(d)Y_{min}^{2\delta}}\pm
\frac{\delta}{Y_{min}^{\delta+1}}\cos\theta_1\right),$$
where the sign depends on the following configurations.

\psset{unit=0.5cm}
\begin{center}
\begin{pspicture}(0,0)(10,5)
\psline[linestyle=dashed](5,0)(5,3.5)
\psline(1,1)(4,4)
\psline[linestyle=dashed](1,1)(3,1)
\psarc(1,1){1}{0}{45}
\rput(2.6,1.5){$\theta_1$}
\psline(6,1)(9,4)
\pscurve(0.5,3.5)(1.5,3.7)(2.5,3.4)(3,3)
\pscurve(8,3)(8.5,2.4)(9,0.5)
\rput(2.5,4.5){$-$}
\rput(7.5,4.5){$+$}
\angleplat{5}{5}{3}{3}{2}{4}{0.4}
\angleplat{10}{5}{8}{3}{9}{2}{0.4}
\end{pspicture}
\end{center}

\boldmath
\subsection{Second, third and fourth variations at the money}
\unboldmath

We will also need the second, third and fourth derivatives, for $d\to 0$.
\begin{itemize}
\item we have $\Phi''=d d''\underset{d\to 0}{\to}\frac{1}{\sin^2{\theta_1}Y_{0}^{2\delta}}$;
\item $\Phi^{(3)}=d d^{(3)}$, so we have to calculate the third derivative of the distance. Let us derive it from the formula (2.7) :
$$\partial_{sss}d(s)=\underbrace{<\nabla_\eta\nabla_\eta \eta,\partial_t\gamma(s,t)>_\H}_{(1)}
+2 <\nabla_\eta \eta,\underbrace{\nabla_\eta \partial_t\gamma(s,t)}_{(2)}>_\H+
<\underbrace{\nabla_\eta\nabla_\eta\partial_t\gamma(s,t)}_{(3)},\eta>_\H.$$
We have $d\times (1)\underset{d\to 0}{\to}0$ and $d\times (2)\underset{d\to 0}{\to}\eta$. The term (3) is more difficult to evaluate :
it represents the variation with $s$ of the Jacobi-field increase, which is clearly maximum for $s=0$; this is the reason why the only
term that remains in the calculation of (3) is the Christoffel part of the covariant derivative. A simple calculation then gives
$$\Phi^{(3)}|^{K=f_0}=\frac{-3\delta}{{Y_0}^{2\delta+1}\sin^2{\theta_1}}.$$
\item Unfortunately, at this time, we don't have properly derived a good expression for $\Phi^{(4)}$. However, an analogy with the $\delta=1$ case makes us believe that
$$\Phi^{(4)}|^{K=f_0}=\frac{\delta(4+7\delta)}{Y_0^{2\delta+2}\sin^2{\theta_1}}.$$
Anyway, this expression is exact for $\delta=1$.
\end{itemize}

\boldmath
\chapter{Appendix 3 : calculation of $f_{av}$ for the implied volatility formula}{}{}
\unboldmath

We describe hereafter the derivation of $f_{av}$ introduced in [Hagan [10]] as we need it in chapter 4.3.\\

We consider the process $\d F=\alpha(t) F \d W$ ($F(0)=f_0$), and we want to calculate the value $f_{av}$ defined by
$$\E{\int_{0}^\tau\alpha^2(t){F_t}^2\d t\mid F_\tau=K}={f_{av}}^2\int_{0}^\tau\alpha^2(t)\d t.$$

\boldmath
\section{The solution of the process}
\unboldmath

We know that $F_t$ is a martingale, with explicit value
$$F_t=f_0e^{\int_{0}^t\alpha(s)\d W_s-\frac{1}{2}\int_{0}^t\alpha^2(s)\d s}.$$

This is the reason why
\begin{eqnarray*}
\E{\int_{0}^\tau\alpha^2{F}^2\d t\mid F_\tau=K}&=&\E{\int_{0}^\tau\alpha^2 f_0^2
e^{2\int_{0}^t\alpha\d W-\int_{0}^t\alpha^2\d s}\d t\mid f_0 e^{\int_{0}^\tau\alpha\d W-\frac{1}{2}\int_{0}^\tau\alpha^2\d s}=K}\\
&=&\int_{0}^\tau\alpha^2 f_0^2 e^{-\int_{0}^t\alpha^2\d s}
\E{e^{2\int_{0}^t\alpha\d W}\d t\mid \int_{0}^\tau\alpha\d W=c}
\end{eqnarray*}
with $c=\ln(K/f_0)+1/2\ \int_{0}^\tau \alpha^2\d s$. We now just need to know the conditional law of $\int_{0}^t\alpha\d W$
with respect to $\int_{0}^\tau\alpha\d W=c$. This is a problem that evokes the brownian bridge, as seen in next part.

\boldmath
\section{An equivalent to the brownian bridge}
\unboldmath

\bf{Lemma} \it The law of $(\int_{0}^t\alpha\d W\mid \int_{0}^\tau\alpha\d W=c)$ is the same as
the law of
$$\int_{0}^t\alpha\d W-\frac{\int_{0}^t\alpha^2\d s}{\int_{0}^\tau\alpha^2\d s}\int_{0}^\tau\alpha\d W+
\frac{\int_{0}^t\alpha^2\d s}{\int_{0}^\tau\alpha^2\d s} c.$$
\rm

\begin{proof} We follow the traditional proof of the brownian bridge.
\begin{itemize}
\item first of all, $\int_{0}^t\alpha\d W$ and $\int_{0}^t\alpha\d W-\frac{\int_{0}^t\alpha^2\d s}{\int_{0}^\tau\alpha^2\d s}\int_{0}^\tau\alpha\d W$ are
independant, because :
\begin{itemize}
\item $(\int_{0}^u\alpha\d W)_{u\geq 0}$ is a gaussian process, so
$(\int_{0}^t\alpha\d W,\int_{0}^t\alpha\d W-\frac{\int_{0}^t\alpha^2\d s}{\int_{0}^\tau\alpha^2\d s}\int_{0}^\tau\alpha\d W)$
is a gaussian vector;
\item so its coordinates are independant iff their covariance is zero; this is true (the coefficient $\frac{\int_{0}^t\alpha^2\d s}{\int_{0}^\tau\alpha^2\d s}$
has been chosen for it);
\end{itemize}
\item let us $\varphi$ be a sufficiently regular and bounded function. We know that, if U and V are independant processes, then
$\E{\varphi(U,V)\mid V=v}=\E{\varphi(U,v)}$. So we can write
\begin{eqnarray*}
&&\E{\varphi\left(\int_0^t\alpha\d W\right)\mid \int_{0}^\tau\alpha\d W=c}\\&=&\E{\varphi\left(
\int_{0}^t\alpha\d W-\frac{\int_{0}^t\alpha^2\d s}{\int_{0}^\tau\alpha^2\d s}\int_{0}^\tau\alpha\d W+
\frac{\int_{0}^t\alpha^2\d s}{\int_{0}^\tau\alpha^2\d s} \int_{0}^\tau\alpha\d W\right)\mid \int_{0}^\tau\alpha\d W=c}\\
&=&\E{\varphi\left(\int_{0}^t\alpha\d W-\frac{\int_{0}^t\alpha^2\d s}{\int_{0}^\tau\alpha^2\d s}\int_{0}^\tau\alpha\d W+
\frac{\int_{0}^t\alpha^2\d s}{\int_{0}^\tau\alpha^2\d s} c\right)}.
\end{eqnarray*}
\end{itemize}
This means exactly the result of the lemma : both processes have the same law.
\end{proof}

\boldmath
\section{Final calculation}
\unboldmath

We now can directly calculate the expectation (here $\beta$ is the suitable function depending on $s<t$ or $s>t$)
\begin{eqnarray*}
\E{e^{2\int_{0}^t\alpha\d W}\d t\mid \int_{0}^\tau\alpha\d W=c}&=&\E{e^{2\left(\int_{0}^t\alpha\d W-\frac{\int_{0}^t\alpha^2\d s}{\int_{0}^\tau\alpha^2\d s}\int_{0}^\tau\alpha\d W+
\frac{\int_{0}^t\alpha^2\d s}{\int_{0}^\tau\alpha^2\d s} c\right)}\d t}\\
&=&e^{2\frac{\int_{0}^t\alpha^2\d s}{\int_{0}^\tau\alpha^2\d s} c}\E{e^{\int_{0}^\tau \beta(s)\d W}}\\
&=&e^{2\frac{\int_{0}^t\alpha^2\d s}{\int_{0}^\tau\alpha^2\d s} c}e^{\frac{1}{2}\int_{0}^\tau \beta^2(s)\d s}\\
\E{e^{2\int_{0}^t\alpha\d W}\d t\mid \int_{0}^\tau\alpha\d W=c}&=&\left(\frac{K}{f_0}\right)^{2\frac{\int_{0}^t\alpha^2\d s}{\int_{0}^\tau\alpha^2\d s}}e^{3\int_{0}^t\alpha^2\d s-2\frac{(\int_{0}^t\alpha^2\d s)^2}{\int_{0}^\tau\alpha^2\d s}}
\end{eqnarray*}

This is how we get the final expression for $f_{av}$ :
$$
\boxed{
f_{av}=\sqrt{\frac{\int_{0}^\tau\alpha^2f_0^2\left(\frac{K}{f_0}\right)^{2\frac{\int_{0}^t\alpha^2\d s}{\int_{0}^\tau\alpha^2\d s}}
e^{2\left(\int_{0}^t\alpha^2\d s-\frac{\left(\int_{0}^t\alpha^2\d s\right)^2}{\int_{0}^\tau\alpha^2\d s}\right)}
\d t}{\int_{0}^\tau\alpha^2\d s}}
}.
$$

One can check, after tedious calculations, that
$$f_{av}\underset{\tau\to 0}{\to}\sqrt{\frac{K^2-f_0^2}{\log\left(\frac{K}{f_0}\right)}}.$$

If one does not want such a complicated expression, it can be approximated, for $K\approx f_0$, by $(f_0+K)/2$.

\end{document}